\documentclass[twocolumn,showpacs]{revtex4}
\usepackage{graphicx,amssymb}  % Include figure files and symbols for figure caption
\begin{document}
\verb| |\\  % Seems to fix some latex bug

\title{In-Medium $\rho^0$ Spectral Function Study via the
  $^2$H, $^3$He, $^{12}$C$(\gamma,\pi^+\pi^-)$ Reaction}

\author{G.M.~Huber}
\author{G.J.~Lolos}
\author{Z.~Papandreou}
\author{A.~Shinozaki}
\author{E.J.~Brash}
\author{M.~Iurescu}
\affiliation{Department of Physics, University of Regina, Regina, SK, S4S 0A2, 
Canada}

\author{G.~Garino}
\author{K.~Maruyama}
\affiliation{Institute for Nuclear Study, University of Tokyo, Tanashi, Tokyo 
188, Japan}

\author{K.~Maeda}
\author{T.~Suda}
\author{A.~Toyofuku}
\affiliation{Department of Physics, Tohoku University, Sendai 980, Japan}

\author{B.K.~Jennings}
\affiliation{TRIUMF, Vancouver, BC, V6T 2A3, Canada}

\author{A.~Sasaki}
\affiliation{College of General Education, Akita University, Akita 010, Japan}

\author{H.~Yamashita}
\affiliation{Department of Applied Physics, Tokyo University of Agriculture 
and Technology, Koganei, Tokyo 184, Japan}

\author{(The TAGX Collaboration)}
\noaffiliation

\date{\today}

\begin{abstract}
We report a helicity analysis of sub-threshold $\rho^0$ production on $^2$H,
$^3$He and $^{12}$C at low photo-production energies.  The results are
indicative of a large longitudinal $\rho^0$ polarization ($l=1$, $m=0$) and are
consistent with a strong helicity-flip production mechanism.  This signature is
used to extract in-medium $\rho^0_L$ invariant mass distributions for all three
nuclei in a manner which is less model-dependent than previous measurements.
The results are compared to kinematic and phenomenological models of the
$\rho^0$ spectral function.  The $^2$H and $^3$He data distributions support
the role of $N^*(1520)$ excitation in shaping the in-medium $\rho^0_L$
invariant mass distribution, while the $^{12}$C distributions are consistent
with quasi-free $\rho^0_L$ production.  The data support an in-medium
modification of the $\rho^0_L$ invariant mass distribution.
\end{abstract}


\maketitle

\section{Introduction}

Of all particles, the $\rho$-meson has received the most attention with regard to
medium modifications.  Since the $\rho^0$ carries the quantum numbers of the
conserved vector current, its properties are related to chiral symmetry, and
can be investigated with a variety of models.  Most models predict a reduction
of the renormalized vector meson mass in the nuclear medium.  A review of the
field, up to 1999, is given in Ref. \cite{Li99}.  One of the first models was
``Brown and Rho scaling'' \cite{Br91}, in which the in-medium $\rho^0$ mass is
rescaled according to the relation
\begin{displaymath}
\frac{m^*_{\sigma}}{m_{\sigma}}\approx
\frac{m^*_N}{m_N}\approx\frac{m^*_{\rho}}{m_{\rho}}\approx
\frac{m^*_{\omega}}{m_{\omega}}\approx\frac{f^*_{\pi}}{f_{\pi}}.
\end{displaymath}
This relation is based on chiral symmetry and scale invariance, and predicts
that the mass of the $\rho^0$ should drop by $\sim 15\%$ from its free mass
value at standard nuclear density.  This is supported by Lattice QCD
calculations \cite{Ji95}, which suggest that chiral symmetry will be fully
restored at $T_c\ge 150$ MeV and/or $\rho_c\ge 5$ $\rho_{nuc}$.

In addition to mass rescaling, the in-medium shape of the $\rho^0$ may be
changed by resonant interactions.  One well-known model of these interactions
is that of Rapp, Chanfray, and Wambach \cite{Ra97}.  Here, the strong coupling
of the $\rho^0$ with $\pi^+\pi^-$ states in the nuclear medium, and
$\rho^0$-baryon scattering, lead to a series of ``rhosobar'' excitations.
This results in an enhancement of the low invariant mass portion of the
$\rho^0$ spectral function.  This model has been applied to the high
temperature and high density regime of high energy heavy ion collisions with
great success.  This picture has been refined and extended by other groups in
recent years, such as the relativistic-model calculation by Post, Leupold and
Mosel \cite{Po01}.  These and other models will be discussed in more detail
later in the paper.

Experimental evidence for in-medium $\rho^0$ mass modification has been
widespread, but all interpretations suffer from significant model
uncertainties.  For example, CERN dilepton production data from S+Au and S+W
collisions at 200 GeV/u yield a significant enhancement at low $m_{ee}$
\cite{CERES}, indicative of a density-dependent $\rho^0$ mass reduction which
is consistent with chiral symmetry restoration as well as with $\rho^0$-medium
rescattering.  An IUCF $^{28}$Si$(\vec{p},\vec{p'})^{28}$Si polarization
transfer experiment found an effective isovector $NN$ interaction strength
consistent with $m_{\rho}\approx 615$ MeV/c$^2$ \cite{St97}, and a KEK 12 GeV
$p+A$ collision experiment yielded $e^+e^-$ spectra which indicate a
significant enhancement for Cu which is not present for $^{12}$C, consistent
with $\approx 200$ MeV/c$^2$ $\rho/\omega$ mass shift \cite{Oz01}, but these
both suffer ambiguities in interpretation due to the use of a proton probe.
Frascati $\gamma A$ total photo-absorption cross sections on C, Al, Cu, Sn, Pb
at 0.5-2.6 GeV are smaller than expected, indicative of increased shadowing.
These results are best explained in terms of a reduced in-medium mass,
$m_{\rho} \sim 610-710$ MeV/c$^2$, which increases the coherence length
$\lambda_{\rho}=2k/m_{\rho}^2$ \cite{bianchi}.  However, due to the inclusive
nature of the experiment, this can only be considered an indirect observation
of in-medium $\rho^0$ properties.

Finally, the TAGX Collaboration has investigated the $\rm
^3He(\gamma,\pi^+\pi^-)$ reaction in the sub-threshold region.  In this case,
diffractive $\rho^0$ production is suppressed, as the reaction must utilize
Fermi momentum to produce a $\rho^0$.  The resulting $\rho^0$'s are produced with
low boost with respect to the nuclear medium, so that they decay within 1 fm of
their production point (i.e. still within even a small nucleus).  These two
properties of the sub-threshold region reaction enhance the nuclear-medium
effect.  The effects of $\pi N$ final state interactions (FSI) are minimized by
the choice of a small nucleus ($^3$He).  An energy-dependent mass reduction
$m^*_{\rho}\approx 640-680$ MeV/c$^2$ was found for $800< E_{\gamma}\leq 1120$
MeV \cite{Lo98,Ka99,Ma00}, with evidence for even lower $\rho^0$ mass at lower
photon energy \cite{Hu98}.  Of these, Ref. \cite{Ma00} used an analysis
technique significantly different than Refs. \cite{Lo98,Ka99}, and yielded
nearly the same $m_{\rho}^*$ value.  A limitation of these results is the
model-dependent separation of the $\rho^0$ channel from other processes leading
to the $\pi^+\pi^-$ state.  Experimental data which can be interpreted in a
less model-dependent fashion are highly desirable, and this is the goal of the
study reported here.

A technique which may allow a cleaner separation of the
$\rho^0\rightarrow\pi^+\pi^-$ contribution from competing processes has
recently been reported by the TAGX Collaboration \cite{Lo02}.  There, the
$\rho^0$ decay angular distribution is reconstructed for sub-threshold $^2$H and
$^{12}$C$(\gamma,\pi^+\pi^-)$ reaction data, and it is found that when
kinematic cuts which enhance the relative population of
$\rho^0\rightarrow\pi^+\pi^-$ decay compared to competing processes, such as
$\Delta\pi$ production, are applied to the data, a strong
cos$^2\theta^*_{\pi^+}$ distribution results.  The only mechanism compatible
with the features of the data is the decay of longitudinally polarized
$\rho^0_L$, consistent with a strong helicity-flip mechanism of $\rho^0$
production.  This result forms the motivation for this work.  Such
helicity-flip amplitudes are interesting, because they may be related to the
$\rho^0$ production mechanism and so assist in such investigations.  Here, the
emphasis is placed on the use of the longitudinal polarization of the
$\rho^0_L$ to isolate its contribution to the $\pi^+\pi^-$ data, with the goal
being to extract the $\rho^0_L$ in-medium line shape and compare it to various
phenomenological models.  A preliminary analysis using this technique was
presented in Ref. \cite{Hu02} and this work presents our final results and
conclusions for $^2$H, $^3$He and $^{12}$C.

The deuteron comprises an important element in the investigation of medium
modifications.  Its low Fermi momentum and nuclear matter density reduce the
probability of medium modifications and/or nucleonic effects.  However, the low
binding energy and the better defined final states for Monte Carlo (MC)
simulations also make the deuteron an attractive test case for comparisons with
data and with the free $\rho^0$ line shape.  $^{12}$C is a ``benchmark'' test
of nuclear modifications, because of its combination of nuclear size and
density.  Unlike the simpler targets, the number of nucleons present in
$^{12}$C raise potential complications due to FSI, including absorption.  The
effects of such FSI will be investigated in detail later in this paper.  $^3$He
may be expected to represent the ``nuclear'' aspects of $^{12}$C with the
simplicity of $^2$H, as far as FSI and pion absorption corrections are
concerned.  The low relative $\rho^0$-N and $\rho^0-^3$He momenta of this
experiment largely compensate for the short nuclear radius disadvantage,
compared to more massive nuclei, by increasing the probability of $\rho^0$
decay within the nuclear volume \cite{Ka99}.

This paper is organized in eight sections.  In Sec. II, the experiment setup
and analysis procedure are reviewed.  Sec. III discusses the MC simulations and
presents representative characteristics of the reaction channels considered.
Sec. IV compares the data distributions with the MC simulation results, with
particular emphasis on the effects of the kinematic cuts on the observed
helicity angle distributions.  In Sec. V, the helicity analysis is used to
quantify and subtract the proportion of non-$\rho^0_L$ events surviving the
kinematic cuts, yielding the experimental in-medium $\rho^0_L$ invariant mass
distributions.  Sec. VI compares the experimental distributions to a series of
model calculations, and Sec. VII compares the helicity angle analysis result
to the previous TAGX Collaboration results.  Finally, in Sec. VIII the
discussion and conclusions are presented.

\section{Experiment and Data Analysis}

The analysis presented here is based on data taken in three separate
running periods \cite{Wa97,Ka99,Lo02}.  Together, they comprise data with
$^2$H, $^3$He and $^{12}$C targets.  The $(\gamma,\pi^+\pi^-)$ experiments were
carried out using the tagged photon beam of the 1.3-GeV Electron Synchrotron
(ES) at the Institute of Nuclear Study at Tokyo (INS) and the TAGX magnetic
spectrometer.

\subsection{Tagged photon beam}

\begin{figure}[hbtp!]
\includegraphics[width=8.5cm]{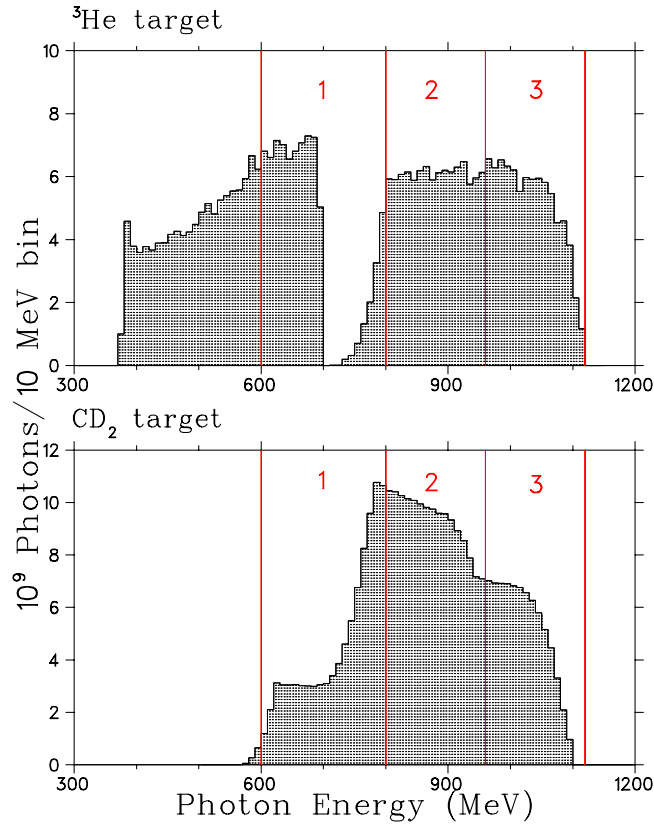}
\caption{\label{fig:egam_spec}
Tagged photon energy distributions used in this work.  Data were
  divided into three $E_{\gamma}$ bins of 600-800 MeV, 800-960 MeV, and
  960-1120 MeV.}
\end{figure}

The photon beam is produced utilizing the 1.3 GeV Tokyo Electron Synchrotron
with a duty factor of $\sim$10\% \cite{Yoshida,Mar96}.  Figure
\ref{fig:egam_spec} displays the tagged photon energy distributions used in
this work.  As the data for the different targets were obtained at different
times, and under slightly different experimental conditions, the shapes of the
resulting tagged photon energy spectra are significantly different for the
$^3$He and CD$_2$ targets.  The $^3$He target data are from Refs. \cite{Ka99}
and \cite{Wa97}, which used ES nominal energies of 800 and 1220 MeV,
respectively.  The CD$_2$ target data were taken with three ES energies of
1040, 1180 and 1200 MeV, with overlapping tagged photon energy distributions.

The data were divided into three tagged photon energy bins of 600-800 MeV,
800-960 MeV, and 960-1120 MeV.  The choice of the limits of these bins was
motivated by the photon energy thresholds for $\rho^0$ production via the
quasi-free and non-quasi-free mechanisms in nuclei.  On $^1$H, the threshold
for production of $\rho^0$ with mass one $\sigma$ lower than the nominal 770
MeV/c$^2$ (i.e. 705 MeV/c$^2$), is 971 MeV.  The division between the mid- and
high-energy photon bins is placed near here, and so the highest photon energy
bin is expected to receive a significant contribution from quasi-free $\rho^0$
production on all three targets.  On $^{12}$C, the corresponding
photo-production threshold is 723 MeV, while for the $\rho^0$ centroid it is 797
MeV.  This motivates the division between the low- and mid-energy bins to be
800 MeV.  In this case, the middle photon energy bin may contain contributions
from the low-mass components of the $\rho^0$ (more than one $\sigma$ below
mean) in a quasi-free production mechanism, as well as nominal $\rho^0$
production from the nucleus as a whole.  The mid-energy bin also has the best
photon luminosity and event statistics for all three nuclear targets.  The
lowest photon energy bin is deeply sub-threshold and so will be the most
sensitive to low-mass components of the $\rho^0$ in a nuclear environment.
However, at these low energies the $\pi^+\pi^-$ production cross-section is
small, and any interpretation of the data will be complicated by significant
phase-space restrictions and non-$\rho^0$ contributions.

\subsection{Targets}

The $^3$He data were obtained with the use of a cryogenic liquid target of 5 cm
diameter and 0.0786 $\rm g/cm^3$ density \cite{tgt3he}.  This necessitated an
empty-target background subtraction procedure, explained in Ref.~\cite{Ka99}.
The $^2$H and $^{12}$C data were obtained together, via the use of a solid
deuterated polyethylene and research grade graphite target assembly.  
The graphite target (0.069 g/cm$^2$) was positioned at the nominal center of
the TAGX magnetic field, with two deuterated polyethylene targets (total
thickness 0.618 $\rm g/cm^2$) flanking it upstream (-16.8 mm) and downstream
(+15.8 mm) relative to the photon beam.  The $^2$H data were obtained by
subtraction of the graphite target data from the CD$_2$ target data, via the
procedure discussed in section \ref{sec:subtract}.

\subsection{TAGX spectrometer}

\begin{figure}[hbtp!]
\includegraphics[width=8.5cm]{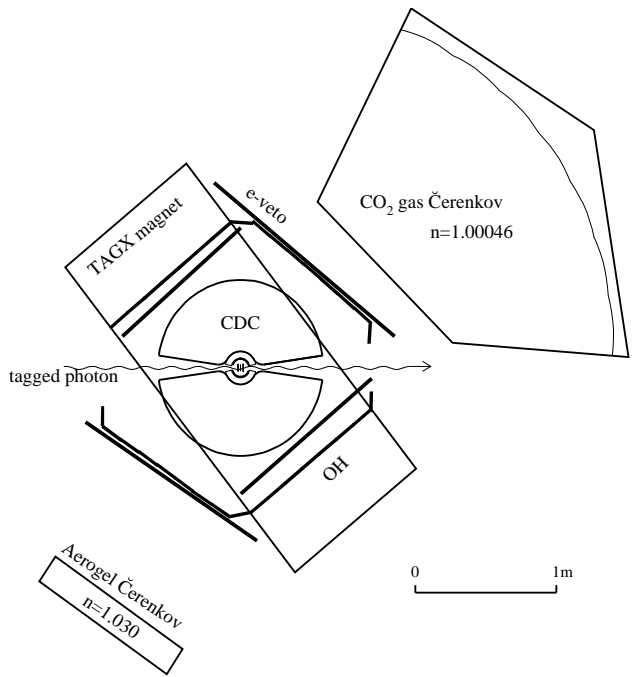}
\caption {\label{fig:TAGX} 
Top view of the TAGX spectrometer elements in place during the $^2$H
and $^{12}$C data taking.}
\end{figure}

The TAGX spectrometer has an acceptance of $\pi$ sr for charged particles.  It
consists of a dipole magnet ($\sim$5 kG), drift chambers for tracking, and
plastic scintillation counters for time-of-flight (TOF) and trigger
information.  Fig.~\ref{fig:TAGX} displays the layout of the TAGX system for
the $^2$H and $^{12}$C measurements.  A brief description of the detectors
follows.  An extended description of the TAGX spectrometer is found in
Ref.~\cite{Mar96}.

Directly surrounding the target is the inner hodoscope (IH), made of two sets
of six scintillator counters, one on each side of the beam.  The IH is used in
the trigger, as well as in measuring the time of flight (TOF) of the outgoing
particles \cite{Mar96,Gar97}.

Next are the two straw tube drift chambers (SDC), located on opposite sides of
the beam. Their operation greatly improves the vertex resolution capability of
the system, which aids in the suppression of experimental background.  The SDC
were designed to preserve the TAGX $\pi$-sr acceptance prior to its
installation, to not impose extensive modifications of the spectrometer, and to
not induce significant energy losses to traversing particles by keeping its
thickness to minimum.  The SDC is explained in detail in Ref.~\cite{Gar97}.

Surrounding the SDC are two semi-circular cylindrical drift chambers (CDC),
subtending angles from 15$^\circ$ to 165$^\circ$ on both sides of the beam in
the horizontal plane, and $\pm$18.3$^\circ$ in the vertical plane.  Together
with the SDC, they are used to determine the planar momentum and in-plane
emission angle of the traversing charged particles, and the vertex position of
the trajectory crossings.

The outer hodoscope (OH) is a set of 33 scintillator elements placed after the
CDC.  Each scintillator is oriented vertically, with PMT's attached at top and 
bottom to determine the track angle relative to the median plane. The two sets of
hodoscopes, IH and OH, measure the TOF of the tracked particles.

Finally, for the $^2$H and $^{12}$C data taking a pressurized CO$_2$
\u{C}erenkov detector was added to further improve the suppression of
beam-related electromagnetic (EM) background.  The gas pressure was slightly
higher than atmosphere, resulting in an electron threshold of 17 MeV/c.  An
aerogel \u{C}erenkov counter with $n=1.030$ was located at backward angle on
beam right.  The pion threshold for this detector was 570 MeV/c.  Because of
their proximity to the TAGX magnetic field, neither detector was used in the
on-line trigger decision.

The remaining components of the TAGX spectrometer are four
155~mm~$\times$~50~mm~$\times$~5~mm scintillator counters, with a primary
function to veto $e^+e^-$ background.  These veto counters are positioned along
the OH arms in the median plane, and eliminate charged-particle tracks
registering within $\Delta z = \pm$ 2.5 mm, mostly affecting forward-focused
$e^+e^-$ pairs produced copiously downstream of the target, but having a small
effect on $\pi^+\pi^-$ events.

\subsection{Event reconstruction}

As our interest lay with $\pi^+\pi^-$ production from the decay of the $\rho^0$
meson, the experiment trigger was set up to record two-charged particle
coincidences on opposite sides of the beam axis.

The reconstruction of the TAGX data is briefly reviewed here; a detailed
account is in Ref.~\cite{Mar96}.  In the analysis, a lab system of co-ordinates
was used, in which the $x$-axis is taken to be along the direction of the photon
beam and the $z$-axis is taken along the TAGX magnetic field.  The trajectory
of a charged particle reconstructed by the SDC+CDC drift chamber system defined
the horizontal component of the momentum, $P_{xy}$, the horizontal tracked
trajectory length, $l_{xy}$, and the tangential direction of the track in the
horizontal plane, $\phi$.  The resulting planar momentum resolution is given by
the relation $\sigma(P_{xy})/P_{xy}\simeq \textnormal{0.090 [GeV$^{-1}$]}
P_{xy} +0.12\times 10^{-3}$ \cite{Sh02}.

The timing difference of the PMT's mounted on both ends of each OH scintillator
gave the the $z$-component of the hit position, which measured the trajectory
length along the $z$-axis ($l_z$), and allowed the magnitude of the momentum in
three dimensions to be calculated.  The out-of-plane trajectory length
resolution was approximately $\sigma(l_z)\sim 1.4$ cm.  The resulting three
dimensional momentum resolution is dominated by $\sigma(P_{xy})/P_{xy}$, above.

The information from each of the two tracks is combined to provide the vertex
position of each event in the $x-y$ plane.  The SDC+CDC $x$ traceback
resolution is 1.0 mm ($\sigma$) \cite{Sh02}.

\subsection{Particle Identification}

\begin{figure}[hbtp!]
\includegraphics[width=8.5cm]{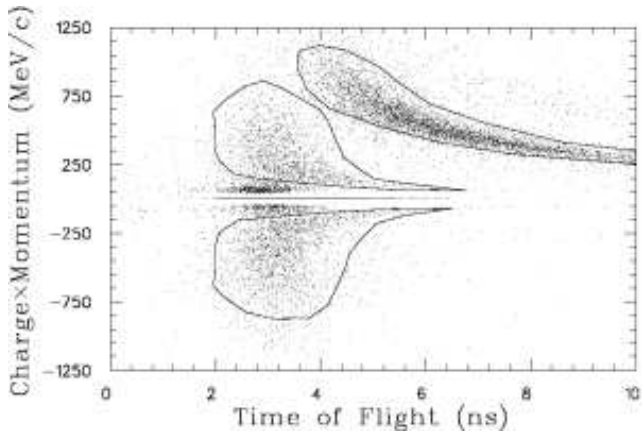}
\caption{\label{fig:pidcut} Data distribution of the product of charge and
momentum versus time of flight through the spectrometer.  The solid lines
indicate the selected regions from which $\pi^+$, $\pi^-$ and proton candidates
were selected for further analysis.  The $e^+e^-$ background is restricted
primarily to the region of small TOF and small momentum.}
\end{figure}

Fig.~\ref{fig:pidcut} shows the observed correlation between the three
dimensional momentum and the TOF between the IH and OH.  Particle
identification boxes were created as shown in the figure to select pions and
protons for further analysis.  This method does not allow $e^+$($e^-$) to be
cleanly separated from $\pi^+$($\pi^-$) because of the limitation set by the
timing resolution.  Therefore, in the case of the CD$_2$ experiment, TDC
information from the forward gas \u{C}erenkov detector was used to additionally
suppress the $e^+(e^-)$ background.  The aerogel \u{C}erenkov detector suffered
from poor gain due to the TAGX magnetic field, and was not used in the
analysis.

Events were only kept if a $\pi^+$ and a $\pi^-$ were detected on opposite
sides of the beam-line, a requirement consistent with the experiment trigger.
Because of the larger probability of one or more protons from the $^{12}$C
target also intercepting the spectrometer, events in which a proton was
detected in addition to the left-right-going $\pi^+\pi^-$ pair were also
accepted for the CD$_2$ experiment, only.  $\pi^+\pi^-p$ events meeting this
condition comprised 6\% of the CD$_2$ target event sample.  ``Three track''
events with a $\pi^+\pi^-$ pair detected on the same side of the beam-line were
excluded from further analysis in all cases.

\subsection{Background Subtraction and Target Separation
\label{sec:subtract}}

For the $^3$He experiment, the cryogenic target was housed in a low-mass target
cell with 185 $\mu$m thick mylar walls, a mylar-aluminum laminate
super-insulator, and a 50 $\mu$m thick aluminum radiation shield \cite{tgt3he}.
This necessitated the use of dedicated empty-target cell data acquisition runs,
for the purpose of background subtraction.  The result of the $^3$He
empty-target cell subtraction is shown in Ref. \cite{Ka99}.

\begin{figure}[hbtp!]
\includegraphics[height=8.5cm,angle=89.9]{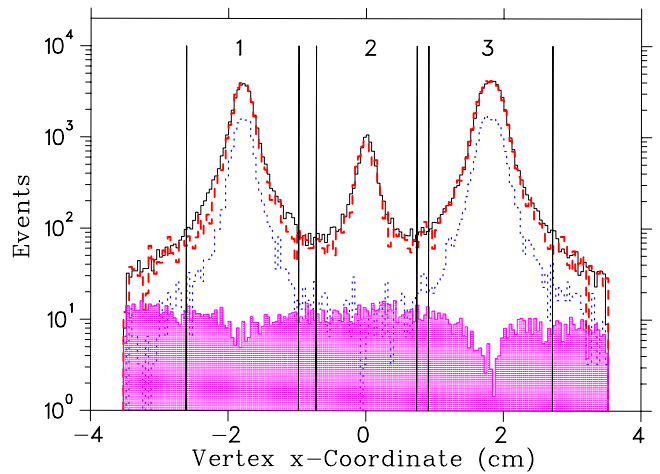}
\caption{\label{fig:3fits} (Color online) Distribution of the $x$-component of
the vertex position for the CD$_2$ experiment data, and compared to a series of
MC simulations taking into account all experiment parameters.  The curves are:
experimental data (solid line), sum of MC simulations (dashed line), simulated
$^2$H contribution (dotted line) and air (shaded region).  To aid the clarity
of the figure, the simulated $^{12}$C contribution is not shown.  The three
indicated regions are the limits of the target cuts used in the analysis.}
\end{figure}

Dedicated empty-target subtraction runs were not necessary for the CD$_2$
experiment, as no target container was used.  In this case, the $x-y$ vertex
position provided the means to identify whether the event originated from the
graphite target, the deuterated polyethylene target, or elsewhere.  Monte Carlo
(MC) simulations incorporating the known target densities and positions, and
taking into account the variation of the traceback resolution upon the
two-track opening angle and the TAGX tracking resolution, were normalized to
the data to provide the target subtraction parameters to extract the $^2$H data
from the CD$_2$ target yield.  Fig. \ref{fig:3fits} compares the $x$-coordinate
distribution of the data to the sum of three MC simulations, in which the
$\pi^+\pi^-$ production was from either $^2$H, $^{12}$C, or the atmosphere
between the solid targets.  The inferred contributions of each source to the
events within each target box shown in Fig. \ref{fig:3fits} are listed in Table
\ref{tab:bgyield}.

\begin{table}[hbtp!]
\caption
{\label{tab:bgyield}Estimated contribution of the various event sources to each
of the target boxes, defined in Fig.~\ref{fig:3fits}, for the CD$_2$
experiment.  The difference between the sum of the contributions and 100\%
indicates the level of agreement between the MC simulations and the
experimental data.}

\begin{tabular}{c|c|ccc|c}
Box \# & Observed Events & $^2$H & $^{12}$C & Air (1 atm) & Sum \\ \hline
   1   & 38487  &  39\%   &  59\% &  1\% &  99\% \\ 
   2   &  9484  &   5\%   &  85\% &  5\% &  95\% \\ 
   3   & 49873  &  38\%   &  61\% &  1\% & 100\% \\ 
\end{tabular}
\end{table}

The three simulations provide an excellent description of the CD$_2$ target regions,
and a poorer but still acceptable description of the $^{nat}$C target region.
In all three cases, the level of agreement is within the 5\% systematic
uncertainty in the MC simulation of the detector, and so it was concluded that
it was not necessary to add an arbitrary background source to describe the
target traceback data.  After application of a correction to account for the
variation in TAGX acceptance with $x$-coordinate (values 1.06 and 0.87 for
Boxes 1 and 3 relative to Box 2), the Box 2 data were subtracted from the Box 1
and Box 3 data to yield the $^2$H event sample.  As air is predominantly N
and O, which will be nearly indistinguishable from $^{12}$C in the physics
analysis, the decision was made to take all of the events within Box 2 as the
$^{12}$C event sample.

\section{Monte Carlo Simulations}

The MC simulations constitute an integral part of the analysis, as they
provide the best means to take into account the effect of the limited
experimental acceptance upon the observed data distributions.  A detailed
comparison of the observed data distributions with those predicted by the MC
simulations will help determine the kinematic regions populated by the
respective reaction channels, and how best to separate their contributions.
Two key observables in this comparison are the lab frame pion-pion opening angle
($\theta_{\pi\pi}$) and the missing mass ($m_{miss}$).  It is necessary for us
to investigate the effects of analysis cuts on these two variables and any
possible inter-dependence between them.  Finally, it is important to show that
the cuts suppress non-$\rho^0$ background in an understood fashion and that
they do not induce distortion in the spectra that can imitate the signature of
a longitudinally polarized $\rho^0_L$.

\subsection{$\pi^+\pi^-$ production channels considered}

\subsubsection{$\rho^0$ production}

We first discuss the $\rho^0$ production generator.  $\rho^0$ production
was assumed to occur via a quasi-free $\gamma N_F \rightarrow \rho^0 N
\rightarrow \pi^+ \pi^- N$ mechanism, where $N_F$ is the participating struck
proton with initial Fermi momentum $p_F$ and the remainder of the nucleus is a
spectator.  Several different $\rho^0$ line-shapes \cite{Est74, Be93, Bu96,
Be98, pdg} were considered in the simulations.  While they all yield very
similar results, all $\rho^0$ simulations shown here use the line-shape of
Benayoun et al. \cite{Be93} unless specified otherwise.  This line-shape is
based on the $l=1$ partial-wave analysis (PWA) of $e^+e^-\rightarrow\pi^+\pi^-$
data from $m_{\pi\pi}=350$ to 1000 MeV/c$^2$, and so is the most authoritative
description of the free $\rho^0$ low mass tail
\footnotemark\footnotetext{Ref. \cite{Be98} by the same authors is based on the
same data, and yields a nearly identical line-shape.}.  The $\rho^0$ then
proceeded to decay into a $\pi^+\pi^-$ pair.  For the simulations assuming
longitudinally polarized $(l=1,\ m=0)$ $\rho^0$ production, the decay
distribution was weighted by cos$^2 \theta_{\pi}$ in the $\rho^0$ rest frame;
for unpolarized $\rho^0$ production the decay distribution was taken to be
isotropic.

\subsubsection{$\sigma^0$ production}

Quasi-free $\sigma^0 \rightarrow \pi^+ \pi^-$ production and its decay is
included as a representative scalar process with Breit-Wigner width and
centroid as listed in Ref. \cite{pdg}\footnotemark\footnotetext{The actual
values used are $m=800$ MeV/c$^2$ and $\Gamma=800$ MeV/c$^2$, which are in the
middle of the range given for the $f_0(400-1200)$.}.  This decay is isotropic in
the $\sigma^0$ rest frame and it is a vital test to investigate whether any
cut-induced $l=1,m=0$-like signatures appear in the MC simulations.

\subsubsection{Baryon resonances}

For the baryon resonances, we follow the work of Refs. \cite{Lo98,Hu98,Ka99},
where a variety of other reaction channels leading to $\pi^+\pi^-$ were
identified and simulated.  Thus, quasi-free reactions leading to the production
of $\Delta^{++}\pi^-$, $\Delta^-\pi^+$, $N^*(1520)\pi^-$,
$N^*(1520)\rightarrow\Delta^{++}\pi^-$ and $N^*(1520)\rightarrow p\rho^0$ were
considered here.  The line-shapes of the baryon resonances were assumed to
be described by Breit-Wigner distributions with centroids and widths taken from
Ref. \cite{pdg}, while the shape of the $\rho^0$ from $N^*(1520)$ decay uses
the PWA shape from \cite{Be93}. In addition, $\Delta^{++}\Delta^-$ production
was simulated in a $\gamma A\rightarrow \Delta\Delta (A-2)$ mechanism, where
$A-2$ was a spectator recoiling in some excited state with the appropriate
Fermi momentum.

As only quasi-free mechanisms have been simulated, $m_{miss}$ cuts must be
placed on both the experimental data and the MC simulated data to exclude
regions populated by more complicated reaction mechanisms involving additional
nucleons or those producing three or more pions.  $\Delta\Delta$ is the lowest
mass mechanism involving more than one target nucleon which leads to the
$\pi^+\pi^-$ final state, and so it serves as an important test case.  Any
kinematic cuts which eliminate the $\Delta\Delta$ process should be even more
effective at eliminating yet more complicated multi-nucleon mechanisms.

The single-nucleon Fermi momentum distributions in the quasi-free simulations
were chosen appropriate to each respective target nucleus.  For $^2$H, a
parametric fit to the $^2$H$(e,e'p)$ data of Bernheim et al. \cite{Bern81} by
A. Saha \cite{Saha} was used.  For $^3$He, a parameterization from a
variational calculation using the Argonne potential \cite{Sch86} for the
2-body breakup $^3$He$(e,e'pp)n$ was used.  For $^{12}$C, the PWIA
parameterizations of J.W. van Orden \cite{VanO} for the $1p_{3/2}$ and
$1s_{1/2}$ shells of $^{16}$O was used, and the recoil nucleus was placed in
the corresponding excited state.  The remaining nucleon or nucleus, as the case
may be, was assumed to be a spectator recoiling with equal and opposite
momentum to the reaction product.

\subsubsection{$\pi N$ final state interactions (FSI)
\label{sec:FSI}}

For all processes leading to $\pi^+\pi^-$ production in the $600 \le E_{\gamma}
\le 1120$ MeV energy range, the emitted pions are in the resonant energy region
and pion re-scattering, as well as absorption, have large cross sections.
Although FSI in the form of absorption may play a role in the extraction of the
final $\rho^0$ invariant mass distribution and will be discussed in more detail
later, re-scattering is of particular concern because helicity signatures are
used to extract the $\rho^0_L$ content, to assign confidence levels and to
define the actual background shapes and strengths to be subtracted from the
data.  While one can argue that for $^2$H and $^3$He re-scattering FSI should be
a small contribution, their effect can not be ignored in the case of $^{12}$C.
For the sub-threshold region considered here, our simulations indicate a mean
$\pi N$ CM momentum of 230 MeV/c, which yields an approximate 50\% probablity
that one of two $\pi^{\pm}$ originating in the center of a uniform density
nucleus with the radius of $^{12}$C will undergo re-scattering.

Therefore, a series of MC simulations were made in which a pion was randomly
selected to scatter with a proton, $\pi+N_F \rightarrow \pi'+N'$, where the
pion initial energy and angle are from the production and decay MC generator,
and the nucleon has appropriate Fermi momentum for $^{12}$C.  The
representative processes for which FSI were studied were quasi-free $\rho^0$
production, $\rho^0$ production via $N^*(1530)\rightarrow \rho^0 p$ decay,
$\Delta^{++}\pi^-$ and $\Delta^{++}\Delta^-$.  The $\pi N$ amplitudes are taken
from the phase shift analysis parameterization of \cite{Ro78}.  

The effect of absorption FSI on the $\rho^0$ line-shape was also considered.  In
this case, a parameterization of the total pion absorption cross-sections of
Ref. \cite{Ash86} and \cite{Jon93} was used to calculate the absorption
probability as the pion traveled from some random point inside a uniform
density sphere with radius of $^{12}$C to the exterior.  Inelastic processes
such as pion induced pion production were not considered since the $m_{miss}$
cut will eliminate them.

In all cases, the simulated particles were transported through the detection
apparatus using GEANT \cite{geant}. The TAGX detector performance evaluated in
the event reconstruction process was utilized to accurately simulate the TAGX
detection efficiency and to reproduce the experimental resolution.  All
experimental thresholds and acceptance cuts were applied, and the simulated
data were analyzed in the same manner as the experimental data.

\subsection{Representative characteristics of the reaction channels}

\begin{figure}[hbtp!]
\includegraphics[width=9.3cm]{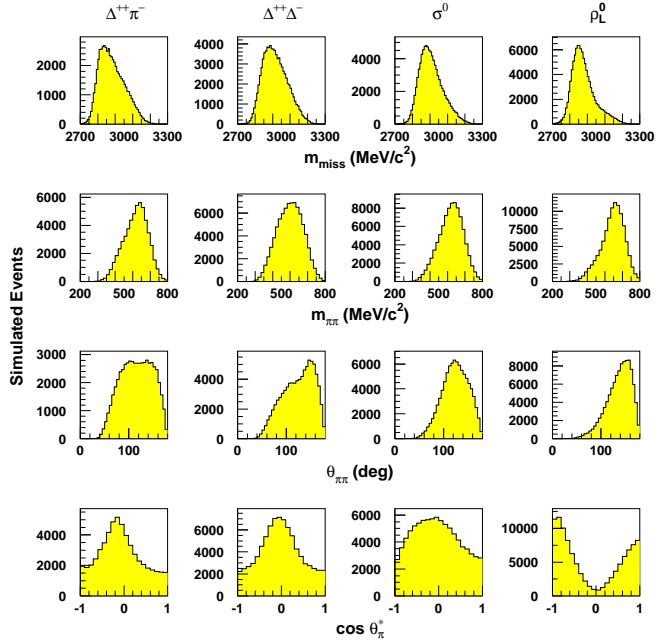}
\caption{\label{fig:nocut4MC} Simulated event distributions for four reactions
on $^3$He, as indicated.  Distributions are as-detected in TAGX, i.e. with full
detection thresholds and experimental acceptance conditions.  All distributions
shown are for the $800<E_{\gamma}\leq 960$ MeV photon energy bin, but no
additional cuts have been applied.  The four simulations shown were each run
for $10^6$ simulated TAGX triggers over the tagged photon range in
Fig. \ref{fig:egam_spec}.}
\end{figure}

Figure \ref{fig:nocut4MC} shows representative kinematic distributions from
four MC simulations with the $^3$He target and for the 800-960 MeV tagged
photon energy range.  The observables plotted are the missing mass $m_{miss}$,
the di-pion invariant mass $m_{\pi\pi}$, the $\pi^+\pi^-$ lab frame opening
angle $\theta_{\pi\pi}$, and, of particular importance to this work, the
helicity angle $\theta^*_{\pi^+}$, calculated in the di-pion rest frame in the
following manner:
\begin{enumerate}
\item{the longitudinal component $p_{\pi^+_{\parallel}}$ of the $\pi^+$
momentum with respect to $\vec{p}_{\pi\pi}$ in the lab frame is formed.}
\item{both $p_+$ and $p_{\pi^+_{\parallel}}$ are transformed to the
$\pi^+\pi^-$ center of mass frame.}
\item{$\theta^*_{\pi^+}$ is then calculated from
\begin{displaymath}
{\rm cos}\theta^*_{\pi^+}=\frac{p_{\pi^+_{\parallel}}}{p_+}.
\end{displaymath}
}
\end{enumerate}

The $^3$He nucleus was chosen for a detailed discussion on the effects
of the cuts on the MC simulations for a number of reasons.  First, $^3$He
exhibits features common to all three nuclei and most of the conclusions drawn
from its analysis apply to the other two.  Second, there are previously
published works on the same target, but based on a different analysis methodology,
that this work can be directly compared to.  Third, the data quality and
statistics (for the MC simulations to be compared to) are good, and this
improves the reliability of any conclusions made.  The $\Delta\pi$ and
$\Delta\Delta$ simulations are selected as representative baryonic non-$\rho^0$
processes and the $\sigma^0$ simulation is a representative mesonic simulation
which will mimic a $\rho^0$ in all respects except for the $l=1,\ m=0$ helicity
signature.

The $\Delta\pi$, $\sigma^0$ and $\rho^0$ simulations make use of a quasi-free
mechanism, as evident by their very similar missing mass distributions,
indicating the escape of a single, energetic nucleon.  The $\Delta\Delta$
simulation involves two energetic nucleons, and so its distribution peaks at
slightly higher missing mass.  For the remainder of the discussion, only events
with missing mass between 2629 and 2919 MeV/c$^2$ were accepted for further
analysis.  The upper limit of range is placed at 110 MeV excitation energy of
the residual nuclear system and so will be referred to as the MM110 missing
mass cut.  It will eliminate all 3$\pi$ production processes.  The value of
the lower limit is not critical, and is placed well below the $^3$He ground
state.

The quasi-free $\sigma^0$ and $\rho^0$ processes have similar lab-frame opening
angle distributions, as each rely on back-to-back $\pi^+\pi^-$ decay in the
meson rest frame.  The $\Delta\pi$ and $\Delta\Delta$ processes have
significantly more strength in the $\theta_{\pi\pi}<100^0$ region, and this is
characteristic of baryonic decay into non-$\rho$ channels.  The cutoff in the
distributions for $\theta_{\pi\pi}<50^0$ is due to the experimental requirement
of two detected tracks on opposite sides of the photon beam-line.

The bottom row of Fig. \ref{fig:nocut4MC} shows the respective helicity angle
distributions of the four processes.  Even though the
$\sigma^0\rightarrow\pi^+\pi^-$ angular distribution is intrinsically isotropic
in cos$\theta^*_{\pi^+}$ due to its $J=0$ nature, once all experimental
considerations are taken into account the resulting distributions are not
isotropic.  The decay of a longitudinally polarized $\rho^0$ is the only process
which produces a cos$^2\theta$-like signature.  The 30\% forward-backward
asymmetry in the simulated $\rho^0_L$ distribution, and the fact that the
distribution maximum is not quite at $\pm 1$, are understood effects of the TAGX
acceptance.

\subsubsection{Investigation of the effects of the $m_{miss}$ and $\theta_{\pi\pi}$ 
cuts
\label{sec:MCcuts}}

The purpose of this section is to observe the effect the missing mass and
opening angle cuts have upon the simulated cos$\theta^*_{\pi^+}$ distributions.
The latter form the basis of the extraction of the $\rho^0_L$ yields and
distribution from the data.

\begin{figure}[hbtp!]
\includegraphics[width=9.3cm]{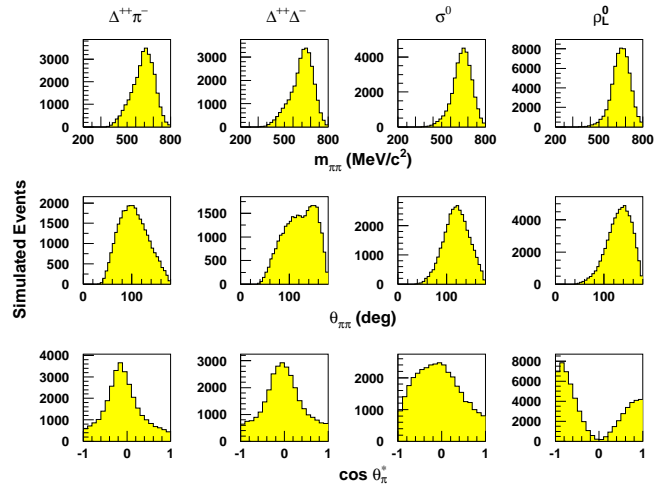}
\caption{\label{fig:mmcut4MC} Simulated event distributions for four reactions
on $^3$He, as indicated.  Distributions are as-detected in TAGX, i.e. with full
detection thresholds and experimental acceptance conditions.  All distributions
shown are with $800<E_{\gamma}\leq 960$ MeV and $2629<m_{miss}\leq 2919$ MeV/c$^2$
(MM110) cuts applied.}
\end{figure}

Fig. \ref{fig:mmcut4MC} shows the same MC simulations as in
Fig. \ref{fig:nocut4MC} after application of the MM110 missing mass cut.  As
expected, the removal of the high missing mass events has depopulated the low
$m_{\pi^+\pi^-}$ region, and so each of the respective invariant mass
distributions have shifted sightly upward in mass.  The opening angle
distribution of the $\Delta\pi$ simulation has shifted to lower values of
$\theta_{\pi\pi}$.  This is a desirable effect, as it will cause the addition
of the opening angle cut to be more effective in reducing the contribution of
this very important process from the data.  The effect of the MM110 cut upon
the other opening angle distributions is much smaller.  Finally, we see that
the effect of this cut upon the helicity angle distributions is small.

\begin{figure}[hbtp!]
\includegraphics[width=8.7cm]{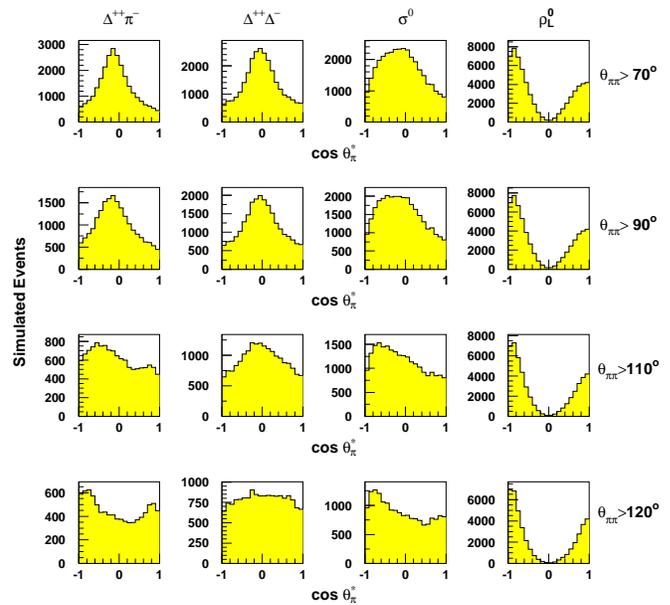}
\caption{\label{fig:thcut4MC} Simulated event distributions for four reactions
on $^3$He, as indicated.  Distributions are as in Fig \ref{fig:mmcut4MC} but
with the addition of an applied opening angle cut, as indicated on the right
side of the figure.  The four simulations shown were each run for $10^6$
simulated TAGX triggers over the tagged photon range in
Fig. \ref{fig:egam_spec}, so the $y$-axis values indicate the survival
probability after the various cuts.}
\end{figure}

Helicity angle distributions for the same four simulations are shown in
Fig. \ref{fig:thcut4MC}, this time with the application of a series of opening
angle cuts of increasing tightness.  The $\theta_{\pi\pi}> 70^o$ cut is
shown for comparison to the previously reported results of Ref. \cite{Ka99}.

One striking observation, by comparing Figs. \ref{fig:nocut4MC} and
\ref{fig:thcut4MC}, is the effectiveness of the two cuts in suppressing
non-$\rho^0$ processes such as $\Delta \pi$ and $\Delta \Delta$.  The
difference in the $y$-axis scale of each figure gives a rough measure of the
effectiveness of the corresponding cuts.  The combination of the MM110 missing
mass cut and the $\theta_{\pi\pi}> 120^o$ cut reduced the $\Delta \pi$,
$\Delta\Delta$ and $\sigma^0$ processes by factors of 6.2, 5.2 and 5.0,
respectively, while reducing $\rho^0_L$ yield by only a factor of 2.4.
These factors are in addition to the inherent selectivity of the TAGX
spectrometer to coplanar $\pi^+\pi^-$ processes such as $\rho^0$ decay.  The
TAGX limited out-of-plane acceptance discriminates against two-step
$\pi^+\pi^-$ production processes and uncorrelated $\pi^+\pi^-$ production when
forming the experiment trigger.  This combination of favorable spectrometer
acceptance and analysis cuts accounts for the experiment's ability to extract a
$\rho^0$ signal with small cross-section out of a comparatively large
background.  Depending on the production processes compared, the relative
enhancement of $\rho^0$ signal can be as much as a factor of ten.

The next observation to be made from Fig. \ref{fig:thcut4MC} is that only the
$\rho^0_L$ simulation produces a strong $p$-wave signature.  It is clear that
an opening angle cut as tight as $\theta_{\pi\pi}> 120^o$ cannot artificially
induce a cos$^2\theta$ distribution in any production process.  In the
$\theta_{\pi\pi}\> 120^o$ row, while the simulated $\Delta^{++}\Delta^-$
distribution is featureless, one may argue that some biasing of the
distributions is evident in the $\Delta^{++} \pi^-$ and $\sigma^0$ channels.
While these two distributions do not resemble a $p$-wave, they do indicate that
some population at the extremes of the cos$\theta^*_{\pi^+}$ distributions can
be accounted for by non-$\rho^0_L$ decay and that care, as well as some
additional criteria, may be necessary to separate $\rho^0$ from non-$\rho^0$
processes.  Nonetheless, for $\theta_{\pi\pi}> 120^o$, no background process
can create the ``edge-to-center'' ratios that $\rho^0_L$ decay exhibits.  In
fact, the population at the center of the distribution is a measure of the
non-$\rho^0_L$ contributions and can be used as an input to further analysis.
We also investigated whether the combination of any two simulated processes can
mimic the observed $\rho^0_L$ distributions.  We found that no combination of
non-$\rho^0$ reactions leading to $\pi^+\pi^-$ production can reproduce
the cos$\theta^*_{\pi^+}$ distribution, and the response to
$\theta_{\pi\pi}$ and $m_{miss}$ cuts, characteristic of $\rho^0_L$
production.  Any admixture of $\Delta^{++}\Delta^-$ together with the other two
background channels will only move the distribution further away from being
$\rho^0_L$-like.

In conclusion, we see that the $m_{miss}$ and $\theta_{\pi\pi}>120^o$ cuts
together provide a good means of enhancing the proportion of $\rho^0$ events in
the data.  This combination of cuts does not appear to induce a false
cos$^2\theta$ distribution to the data, and if used in concert with the
``edge-to-center'' ratio of the distribution, can provide a unique means to
identify the $\rho^0_L$ yield.  While the $\theta_{\pi\pi}>120^o$ cut appears
to be effective, we do not believe it is prudent to cut the data more severely
than this.  Beyond this point, the degraded statistics of the surviving data
and the restricted kinematics caused by the cut become progressively more
important factors.

\subsubsection{Application of cuts to $^2$H and $^{12}$C simulated data}

For the $^2$H target nucleus, the effects of the two cuts on the
cos$\theta^*_{\pi^+}$ distributions for the various reaction processes lead to
conclusions very similar to those presented for $^3$He.  However, given the
more complex nature of $^{12}$C, two specific investigations are worth
exploring in some detail.  One is based on the fact that this nucleus consists
of six protons and six neutrons and charge symmetric background processes may
have differing distributions, due to triggering requirements.  The other is
exploring the effects of FSI, an effect that also has its basis on the larger
number of nucleons present compared to the other two nuclei.

\begin{figure}[hbtp!]
\includegraphics[width=8.7cm]{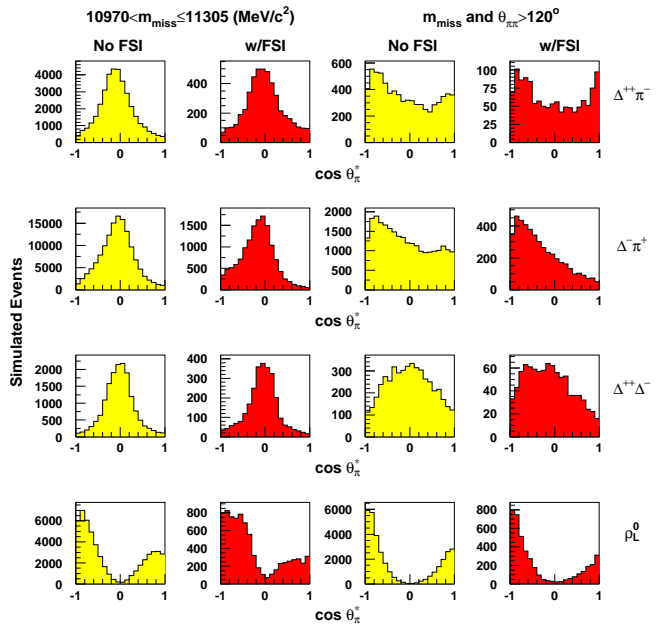}
\caption{\label{fig:cut12CMC880} (Color online) Simulated event distributions
for four quasi-free reactions on $^{12}$C, both with and without the effect of
re-scattering FSI, as indicated.  In the case of the with-FSI simulations, one
of the two pions was randomly selected to undergo $\pi p$ scattering, weighted
according to the phase shift analysis parameterization of Ref. \cite{Ro78}.
All distributions are as-detected in TAGX (i.e. detection thresholds and
experimental acceptance included) and are for $800<E_{\gamma}\leq 960$ MeV and
$10970<m_{miss}\leq 11305$ MeV/c$^2$ (MM130).  In addition, the opening angle
cut $\theta_{\pi\pi}>120^o$ has been applied to the two rightmost columns.  The
four simulations shown were each run for $10^6$ simulated TAGX triggers over
the tagged photon range in Fig. \ref{fig:egam_spec}, so the $y$-axis values
indicate the survival probability after the various cuts.}
\end{figure}

Fig. \ref{fig:cut12CMC880} shows simulations for three non-$\rho^0$ channels
as well as quasi-free $\rho^0$ production, both with and without the effect of
re-scattering FSI.  The upper limit of the missing mass cut applied,
$10970<m_{miss}\leq 11305$ MeV/c$^2$, corresponds to 130 MeV excitation of the
recoil system, and so is termed the MM130 cut.  This higher upper cut value
chosen (as opposed to MM110) is to maximize the data statistics, which
are more limited for $^{12}$C than the other nuclei, while still remaining
below $3\pi$ production.

With only the MM130 cut applied, all non-$\rho^0$ processes exhibit
distributions which peak near the center of the cos$\theta^*_{\pi^+}$ range.
With six protons and six neutrons present, the charge symmetric $\Delta^{++}
\pi^-$ and $\Delta^- \pi^+$ reactions should contribute at approximately the
same rates.  However, the inclusion of re-scattering FSI and the slightly
different TAGX acceptances for $\pi^+$ and $\pi^-$ may induce a difference
between the charge symmetric distributions.  Nonetheless, no artificial
$p$-wave behavior is seen as a result of the MM130 cut.  This cut is also
very effective in suppressing $\Delta\Delta$ processes.

For the non-$\rho^0$ processes, the application of the $\theta_{\pi\pi}>120^o$
cut results in the suppression of yield in the regions between -0.7 and 0.7,
and the remaining population about cos$\theta^*_{\pi^+}=0$ is a measure of the
total surviving background.  All of the simulations in
Fig. \ref{fig:cut12CMC880} were run for the same number of simulated TAGX
triggers, so the poor statistics for the FSI simulations indicate their low
survival probability after the cuts.  The mean cut survival probability of FSI
processes is less than 25\% of that of non-FSI processes, enhancing the expected
relative proportion of non-FSI to FSI events in the $^{12}$C data by a factor
of four.  It is interesting to note that the $\Delta \pi$ channels
exhibit distributions that may be fitted by a function, such as $A+B$
cos$\theta^*_{\pi^+} + C$ cos$^2 \theta^*_{\pi^+}$, that has a
$p$-wave component.  It is also worth noting that the ``skewness'' observed for
the $\Delta^{++} \pi^-$ reaction (forward-backward asymmetry) will give
different values for the three coefficients than that expected for the
$\rho^0_L$ decay process.  These aspects of the analysis will be pursued for
all three nuclei in a later section of this work.

\section{Data distributions and comparisons with MC simulations} 

In this section, the effects of the two cuts, $m_{miss}$ and $\theta_{\pi\pi}$,
on the data distributions are investigated and detailed comparisons with
the MC simulations are made.  The objective is to establish the consistency of
the conclusions derived from the application of the cuts on the data and the
simulations.  In subsequent sections in this paper, this will lead to the
identification of a $\rho^0_L$ component in the experimental data and its
extraction and separation from the non-$\rho^0$ background.

\subsection{The effects of the cuts upon the $^3$He data}

\begin{figure}[hbtp!]
\includegraphics[width=9.3cm]{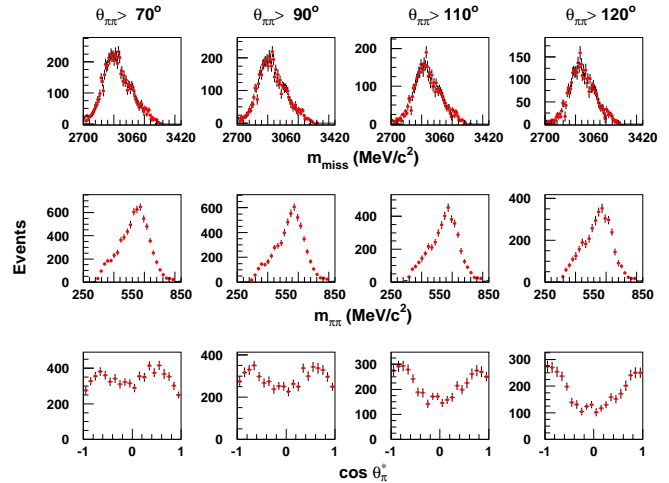}
\caption{\label{fig:3hethcut} $^3$He data distributions for
$800<E_{\gamma}\leq 960$ MeV.  In addition, the indicated opening angle
$\theta_{\pi\pi}$ cuts have been applied to the data.}
\end{figure}

The effect of the $\theta_{\pi\pi}$ cut upon the $^3$He data distributions for
$m_{miss}$ and other variables is shown in Fig. \ref{fig:3hethcut}.  Four
successive $\theta_{\pi\pi}$ cut values of $70^o,\ 90^o,\ 110^o$ and $120^o$
have been applied.  The effect of the cut upon $m_{miss}$ is small.  There is a
shift in the mean of the distribution of approximately 30 MeV/c$^2$ as the cut
is changed from $70^o$ to $120^o$.  The low missing mass region is most
affected, due to the elimination of two pion events with low relative momenta.
The effect on the invariant mass, $m_{\pi\pi}$, is also small.  Neither the
peak nor the shape of the distributions change appreciably for the four values
of $\theta_{\pi\pi}$ cut.  The low mass tail of the distribution, below 500
MeV/c$^2$, is affected a little more greatly, resulting in a slightly more
symmetric shape for the distribution.  Generally, invariant mass distributions
for such multi-channel (inclusive) processes are not sensitive variables on
their own.  This is because they represent phase-space and acceptance
limitations compounded with broad and overlapping resonances.

\begin{figure}[hbtp!]
\includegraphics[width=9.3cm]{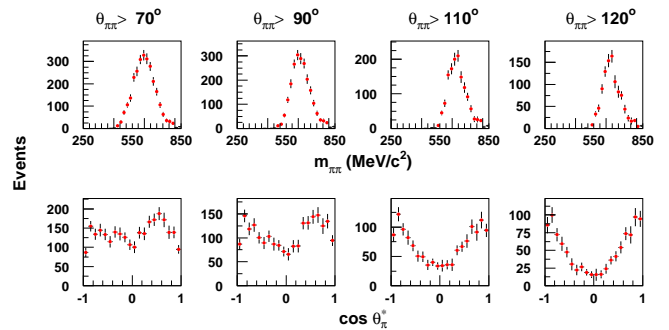}
\caption{\label{fig:3hemmthcut} $^3$He data distributions for
$800<E_{\gamma}\leq 960$ MeV and $2629<m_{miss}\leq 2919$ MeV/c$^2$ (MM110
cut).  In addition, a series of opening angle $\theta_{\pi\pi}$ cuts have
been applied to the data, as indicated.}
\end{figure}

The effect of the MM110 missing mass cut, in addition to the opening angle cut,
is shown in Fig. \ref{fig:3hemmthcut}.  A comparison of
Figs. \ref{fig:3hethcut} and \ref{fig:3hemmthcut} reveals that the application
of the $m_{miss}$ cut has a larger effect on $m_{\pi\pi}$ than the
$\theta_{\pi\pi}$ cut had alone.  By eliminating three pion production and
suppressing the emission of energetic nucleons, the missing mass cut helps to
isolate two energetic pion events adding up to larger invariant mass.  As an
example, the peak of $m_{\pi\pi}$ corresponding to the $120^o$ cut has shifted
from approximately 580 MeV/c$^2$ in Fig. \ref{fig:3hethcut} to approximately
665 MeV/c$^2$ in Fig. \ref{fig:3hemmthcut}.  Similar behavior was observed in
\cite{Lo02} for $^2$H.  These shifts become a significant factor in the
discussion on the $\rho^0_L$ invariant mass distribution and comparisons with
theoretical models in the relevant sections of this work.

The effect of the opening angle cut upon cos$\theta^*_{\pi^+}$ is quite
dramatic.  The effect of the additional MM110 cut has been to make the
resulting cos$^2\theta$ distribution more symmetric.  It is instructive to
compare the corresponding cut plots in Figs. \ref{fig:thcut4MC} and
\ref{fig:3hemmthcut}.  The MC simulations show cos$\theta^*_{\pi^+}$
distributions for all background processes that are radically different from
the data.  Furthermore, no combination of $\Delta\pi,\ \Delta\Delta$ and
$\sigma^0$ processes without $\rho^0_L$ content can reproduce the shape and
change of population of the data in going from the $70^o$ to the $120^o$ cut.
The ``edge-to-center'' ratio is an additional piece of information, as the MC
simulations indicate that the bulk of the background processes are located in
the ``center'' region of the cos$\theta^*_{\pi^+}$ distribution.  Thus, this
ratio is a measure of the $\rho^0_L$ to non-$\rho^0$ contributions to the
observed data yield, and a comparison of Figs. \ref{fig:3hethcut} and
\ref{fig:3hemmthcut} indicates that the effect of the $m_{miss}$ cut has been
to remove additional non-$\rho^0_L$ processes, such as $3\pi$ production.

In conclusion, the two kinematic cuts, applied to both data and MC
simulations, have been shown to be largely uncorrelated, each selecting
different features of the data-set.  They help to establish that the only
self-consistent process identified as the source of the $p$-wave-like
distribution in cos$\theta^*_{\pi^+}$ is the decay of a longitudinally
polarized $\rho^0_L$.  By comparing the `edge-to-center' ratios of the
distributions in Figs. \ref{fig:thcut4MC} and \ref{fig:3hemmthcut}, it is clear
that the non-$\rho^0_L$ background under the $120^o$ cut is minimal.  This is
established by the MC simulations for the $\Delta \pi$, $\Delta \Delta$ and
$\sigma^0$ processes and the small number of data events at
cos$\theta^*_{\pi^+}=0$.  Since this ``pedestal'' is an accumulation of
surviving events from all background processes, it establishes an absolute
measure of the contribution of these processes to the data.

One question that arises as a result of the comparison of the simulations and
the data is that of helicity conserving $\rho^0_T$ or unpolarized $\rho^0$
content in the data and their signatures in the data and simulations.  Both
unpolarized $\rho^0$ and transversely polarized $\rho^0_T$ events, the latter
with a distribution peaking at zero values of cos$\theta_{\pi}$, would add to
the ``pedestal'' at the center of the distribution.  As such, it is clear that
there is little or no evidence for such polarization states of $\rho^0$ in the
data sample.  The argument against $\rho^0_T$ signatures is particularly strong
because such a helicity state would otherwise share all other responses to
$\theta_{\pi\pi}$ and $m_{miss}$ cuts with $\rho^0_L$.  In this case,
increasing the $\theta_{\pi\pi}$ cut limit would result in a central
enhancement, an expectation completely in variance with the data.

\subsection{The effects of the cuts upon the $^2$H data}

\begin{figure}[hbtp!]
\includegraphics[width=9.3cm]{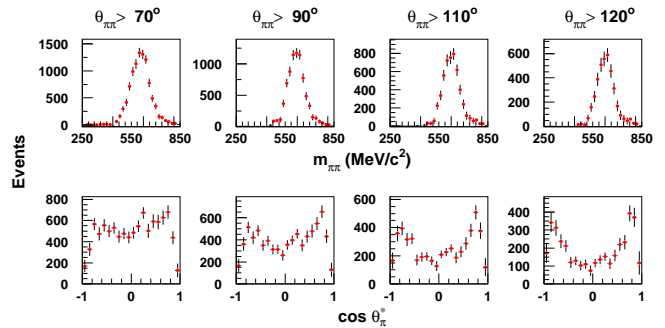}
\caption{\label{fig:dmmthcut} $^2$H data distributions for
$800<E_{\gamma}\leq 960$ MeV and $1745<m_{miss}\leq 2006$ MeV/c$^2$ (MM130
cut).  In addition, a series of opening angle $\theta_{\pi\pi}$ cuts have
been applied to the data, as indicated.}
\end{figure}

The $^2$H data distributions respond to the $\theta_{\pi\pi}$ cuts in a manner
similar to those in Fig. \ref{fig:3hethcut}, except that the $m_{miss}$
distributions show a peak at the deuteron mass with a narrower distribution
than the $^3$He distributions, and the $m_{\pi\pi}$ distributions have slightly
higher centroids, a manifestation of the smaller phase-space available and the
more well-defined final state.  Application of the missing mass MM130 cut
results in the data distributions of Fig. \ref{fig:dmmthcut}.  The effect of
the $m_{miss}$ cut on $m_{\pi\pi}$ is similar to that already shown on $^3$He.
The low invariant mass regions below 500 MeV/c$^2$ are essentially eliminated;
the means of the distributions are approximately 25 MeV/c$^2$ below their
corresponding distributions in Fig. \ref{fig:3hemmthcut}.  This is consistent
with the lower center-of-mass energies available for pion production due to the
lower Fermi momenta of the struck nucleons.

The MM130 cut does not have the same quality of improvement in the
$p$-wave-like behavior of the cos$\theta^*_{\pi^+}$ data as it did for $^3$He.
This is partly due to the fact that the deuterium data are the result of a
subtraction, CD$_2 - ^{12}$C, and partly due to the MM130 cut itself.  This cut
is perhaps too generous for such a loosely bound nucleus, and will be tightened
in subsequent sections of this work.  Nevertheless, the ``edge-to-center''
ratios of cos$\theta^*_{\pi^+}$ distribution for the $120^o$ cut are good.

\subsection{The effects of the cuts upon the $^{12}$C data}

\begin{figure}[hbtp!]
\includegraphics[width=9.3cm]{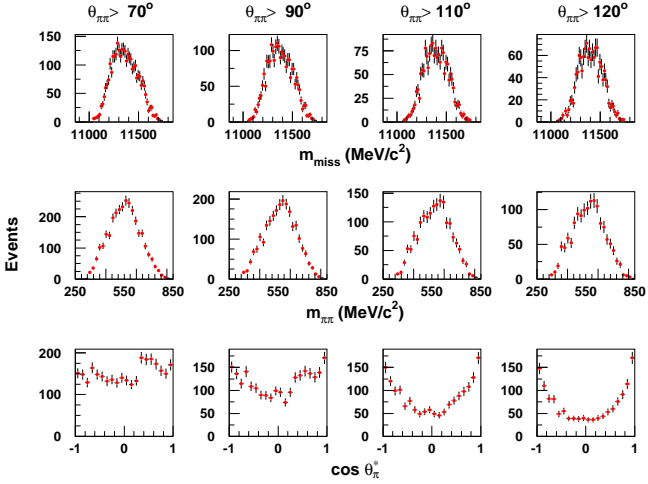}
\caption{\label{fig:cthcut} $^{12}$C data distributions for $800<E_{\gamma}\leq
960$ MeV.  The indicated opening angle $\theta_{\pi\pi}$ cuts have been
applied to the data.}
\end{figure}

Fig. \ref{fig:cthcut} shows the effect of the $\theta_{\pi\pi}$ cuts on the
$^{12}$C data.  The $m_{miss}$ and $m_{\pi\pi}$ distributions are similar to
the two lighter nuclei, but with broader, smoother distributions, reflecting
the larger phase-space available to the pions due to the higher Fermi momentum
in this nucleus.  The overall effect on cos$\theta^*_{\pi^+}$ is also similar
for all three nuclei.  One minor difference observed in this figure is that the
loss of acceptance observed at the extremes of the distribution is even less
pronounced for $^{12}$C compared to $^3$He.  This is consistent with the MC
simulations.  The $p$-wave signature is clearly observed as a result of the
opening angle cut.

\begin{figure}[hbtp!]
\includegraphics[width=9.3cm]{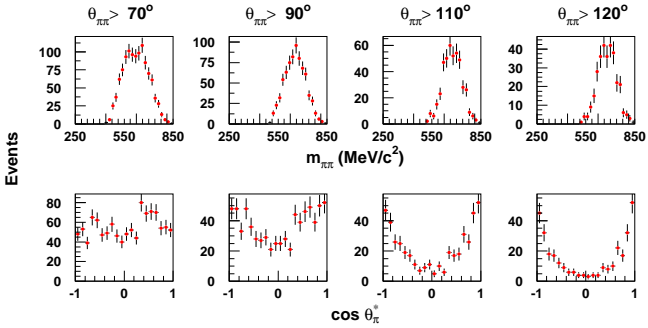}
\caption{\label{fig:cmmthcut} $^{12}$C data distributions for
$800<E_{\gamma}\leq 960$ MeV and $10970<m_{miss}\leq 11305$ MeV/c$^2$ (MM130
cut).  In addition, a series of opening angle $\theta_{\pi\pi}$ cuts have
been applied to the data, as indicated.}
\end{figure}

Application of the MM130 cut results in the distributions in
Fig. \ref{fig:cmmthcut}.  The reduction in the number of surviving events is
dramatic, compared to the other two nuclei.  Application of the same cut on
$^2$H resulted in a survival fraction of 54\% for the $\theta_{\pi\pi} \ge
120^o$ cut.  In the $^3$He case, the surviving fraction is 34\% while in this
case the fraction is 19\%.  This severely restricts the surviving statistical
precision.  However, comparing the cos$\theta^*_{\pi^+}$ distributions for the
$90^o$, $110^o$ and $120^o$ cuts in Figs. \ref{fig:cthcut} and
\ref{fig:cmmthcut}, the MM130 cut reduces the non-$p$-wave background in a very
effective manner.

\subsection{Conclusions from the comparison of the data to the simulations}

\begin{figure}[hbtp!]
\includegraphics[width=9.0cm]{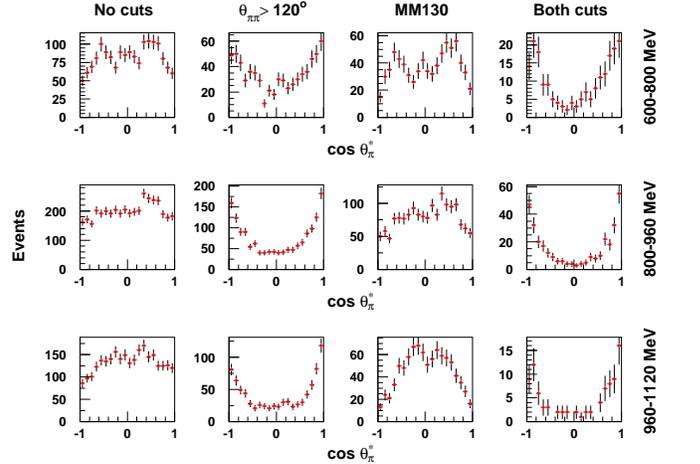}
\caption{\label{fig:c3egamma} Sample $^{12}$C data distributions for the three
  tagged photon energy bins, as indicated.}
\end{figure}

Overall, the effect of the two cuts on all the variables investigated and for
all three different nuclei do not result in any significant differences in the
distributions among the three target nuclei.  It should also be mentioned that
the cos$^*\theta_{\pi^+}$ distributions for the three photon energy bins of
600-800, 800-960 and 960-1120 MeV are also very similar.  A sample plot of the
$^{12}$C data for the full photon energy range is shown in
Fig. \ref{fig:c3egamma}.  Application of the $\theta_{\pi\pi}>120^o$ cut
removes most of the yield near cos$\theta^*_{\pi^+}$=0, resulting in the
$p$-wave-like distributions in the second column.  Application of only the
MM130 cut eliminates events across the distributions, without altering their
shapes. The final column displays the final distributions after the application
of both cuts.  The surviving events are consistent with processes dominated by
$p$-wave-like distributions.  The two-step processes analogous to those
simulated in Fig. \ref{fig:mmcut4MC} cannot be the main contributors, as
they have different responses to the cuts than the foreground channels and the
data respond to the cuts as the simulations predict.  There is little evidence
that background processes dominate the data-set after application of the two
kinematic cuts, each with their own unique effects on both the data and the MC
simulations.  In fact, a direct comparison of Figs. \ref{fig:mmcut4MC} and
\ref{fig:3hemmthcut} indicates that the total background is a small fraction of
the total event sample.  However, the exact contribution of the background in
the cos$\theta^*_{\pi^+}$ distributions needs to be determined in a
quantitative manner, and this is the subject of the next section.

\section{Investigation of non-$\rho^0_L$ background via helicity analysis}

The purpose of this section is to investigate the relative contributions of the
various $\pi^+\pi^-$ production processes to the observed cos$\theta^*_{\pi^+}$
distributions, to subtract the contributions of the background processes in as
model-independent manner as possible, and in so-doing to obtain a better
measure of the observed $\rho^0_L$ distribution versus invariant mass for each
of the target nuclei.  In order to gain more insight into the nature of the
non-$\rho^0_L$ background remaining after the missing mass and opening angle
cuts, both the experiment and the MC-simulated cos$\theta^*_{\pi^+}$
distributions were fit with a function that contains a $p$-wave term, a
$s$-wave term and an interference term.  In this manner, all of the elements of
the distributions, namely a uniform event distribution, a symmetric $p$-wave
distribution about $\theta^*_{\pi^+}=90^o$ and an asymmetric interference term
are represented in the fit.  The fitted function is
\begin{equation}
\label{eqn:abc}
Events = A+B {\rm cos}\theta^*_{\pi^+}+C {\rm cos}^2\theta^*_{\pi^+}.
\end{equation}
This function is not intended to provide a perfect fit to the observed
distributions, but rather to provide a means to quantify the relevant features
of the observed distributions.

If the data show high fractions of $p$-wave content, it means that the
coefficient $A$, which is the ``pedestal'' at cos$\theta^*_{\pi^+}=0$, is very
small.  This can either mean that the background is negligible and the data are
almost purely $\rho^0_L$ in content, or that there may be significant
background, but it is almost purely $p$-wave-like due to the cuts imposed.
None of the background processes in Fig. \ref{fig:thcut4MC} appear to reflect
such a possibility.

In order to extract the in-medium $\rho^0_L$ invariant mass distribution, one
needs to investigate the data and MC simulations as a function of $m_{\pi\pi}$
and possibly extract different confidence levels for different $m_{\pi\pi}$
values.  Both the experimental and MC-simulated data were binned according to
$\pi^+\pi^-$ invariant mass, and a cos$\theta^*_{\pi^+}$ distribution formed
for each $m_{\pi\pi}$ bin.  Each invariant mass bin was 90 MeV/c$^2$ wide,
staggered in 30 MeV/c$^2$ increments.  Thus, the first bin was
360$<m_{\pi\pi}\leq$450 MeV/c$^2$, the second was 390$<m_{\pi\pi}\leq$480
MeV/c$^2$, and so on.  This binning procedure was necessary to ensure
sufficient statistics per bin to allow inferences to be formed, while
preserving sufficient invariant mass resolution to make these inferences
interesting and useful.  The three coefficients $A,\ B,\ C$ were determined for
each bin for both the data and the various MC-simulated processes, and
tabulated according to the mean $m_{\pi\pi}$ value of the respective population
in each bin.  Thus, although the data and simulations used the same binning
scheme, the $m_{\pi\pi}$ tabulated values for each differ slightly according to
the distributions of the populations of each.  Checks were made to ensure that
the resulting parameter distributions versus $m_{\pi\pi}$ were not sensitive to
the bin size and stagger offset used.

If the non-$\rho^0_L$ processes had cos$\theta^*_{\pi^+}$ distributions
described by only the flat $A$ and skewed $B$ terms of the fitting function,
and if the $\rho^0_L$ process had a pure cos$^2\theta^*_{\pi^+}$ distribution
with only the $C$ coefficient being non-zero, then a plot of $C$ versus
invariant mass would give a precise measure of the observed $\rho^0_L$
distribution for each of the target nuclei.  To first order, this is in fact a
good approximation.  However, as we have seen in section \ref{sec:MCcuts}, the
TAGX acceptance modifies the expected $\rho^0_L$ distribution from a pure
cos$^2\theta^*_{\pi^+}$ function, and the background processes contribute
somewhat to $C$ as well as to $A$ and $B$.

If one integrates equation \ref{eqn:abc} from cos$\theta^*_{\pi^+}$=-1 to +1,
one obtains $2A+\frac{2}{3}C$.  The ratio $\frac{C}{3A+C}$ thus provides a
relative measure of the cos$^2\theta^*_{\pi^+}$ component in the observed
distribution.  The skewness coefficient, $B$, does not contribute to the
integral as this term is odd in cos$\theta^*_{\pi^+}$.  However, as we have
seen in Fig. \ref{fig:3hethcut}, certain background processes have
significantly skewed distributions, so the value of $B$ could assist in
the discrimination between $\rho^0_L$ and non-$\rho^0_L$ processes.  This
conclusion was confirmed after further analysis, and so a $p$-wave shape
ratio
\begin{displaymath}
R_{p-wave}=\frac{C}{3|A|+\frac{3}{2}|B|+|C|}
\end{displaymath}
was formed for further study.  This weighs the skewed coefficient as if the
absolute value of the skewness term had been taken in the integral.  The
absolute values in the denominator ensure that cancellation between opposite
sign coefficients cannot artificially boost the value of the ratio.

\subsection{Comparison with data}

\begin{figure}[hbtp!]
\includegraphics[width=8.3cm]{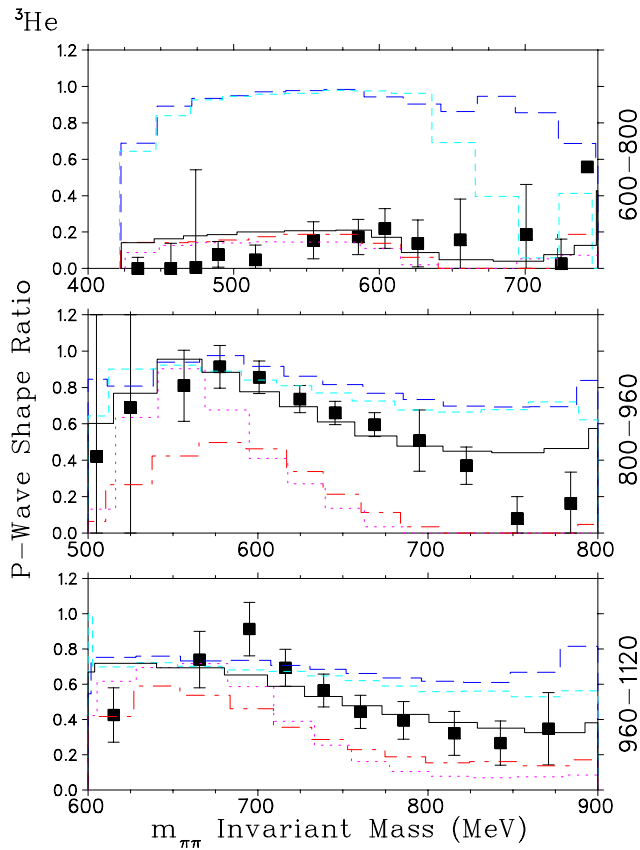}
\caption{\label{fig:comp_mch} (Color online) $p$-wave shape ratios for the
$^3$He data [$\blacksquare$], quasi-free $\rho^0_L$ simulation [long-dash
line], $N^*(1520)\rightarrow p\rho^0$ simulation [short-dash line],
$\Delta^{++}\pi^-$ simulation [dash-dot line] and $\sigma^0$ simulation [dotted
line], for the three tagged photon energy bins, as indicated.  The solid lines
are normalized fits of the simulations to the $R_{p-wave}$ data, as described
in the text.  The regions of largest discrepancy between fit and data are
generally on the tails of the event distributions. All distributions shown are
with $\theta_{\pi\pi}>120^o$ and $2629<m_{miss}\leq 2919$ MeV/c$^2$ (MM110)
cuts applied.}
\end{figure}

Fig. \ref{fig:comp_mch} displays the observed $p$-wave shape ratio for the
$^3$He data and several MC simulations.  The two $\rho^0_L$ producing channels
considered are quasi-free $\rho^0_L$ production and
$N^*(1520)\rightarrow\rho^0N$ decay, while the two ``benchmarks'' for
non-$\rho^0_L$ di-pion production are the quasi-free $\sigma^0$ and $\Delta\pi$
processes.  The elimination of the $\Delta\Delta$ process from further
consideration is based on the observation that it has no $p$-wave-like
structure independent of any cuts applied, and that the proportion of
$\Delta\Delta$ events surviving the cuts is small (see Figs. \ref{fig:thcut4MC}
and \ref{fig:cut12CMC880}).  For the quasi-free $\rho^0_L$ simulation,
$R_{p-wave}$ deviates from 1.0 because of the finite TAGX acceptance, primarily
due to the inability to detect pions emitted between 0 and $15^o$ in the lab
frame.  As expected, the $N^*(1520)\rightarrow\rho^0N$ simulation yields
similar results.  The ratios for the two background processes, however, show
much greater variation, as a result of the $m_{miss}$ and $\theta_{\pi\pi}$
cuts.  For example, at $960<E_{\gamma}\leq 1120$ MeV the $\sigma$ and
$\Delta\pi$ channels display high $p$-wave ratios between $600<m_{\pi\pi}\leq
700$ MeV/c$^2$, but lower ratios in most other regions.

For 800-1120 MeV, the data have $p$-wave ratios which are mid-way between the
$\rho^0_L$-producing and background simulations.  This means that after the
application of the $m_{miss}$ and $\theta_{\pi\pi}$ cuts, the data still
contain some contribution from non-$\rho^0_L$ processes.  The effect of the
background processes is to dilute the $p$-wave shape ratio observed for the
data.  For example, between 800 and 960 MeV tagged photon energy, a less than
10\% contribution of the $\sigma^0$ process to the remaining data would be
sufficient to account for the degradation of the ratio from that of the
pure-$\rho^0_L$ process to that observed for the data.  At 600-800 MeV,
however, the degradation of the observed ratio is more severe, and a more
significant contribution by background processes to the remaining data after
the cuts would be necessary to account for the observed ratio.  For the higher
$E_{\gamma}$ region of 960-1120 MeV, the picture is even more complicated.
Within the data error bars, the $600<m_{\pi\pi}\leq 740$ MeV/c$^2$ region has a
high $p$-wave shape ratio, but at the same time, the background channels
(especially the $\sigma^0$ channel) also have high ratios.  Thus, the level of
confidence in this region is not as high as that of the mid-energy region.

Thus, the confidence in the accurate extraction of the $\rho^0_L$ component,
and therefore its invariant mass distribution, depends on both $E_{\gamma}$ and
$m_{\pi\pi}$.  In the mid-$E_{\gamma}$ range, where the most and highest
quality data have been obtained, the degradation in the observed $p$-wave ratio
appears to be small, and so $\rho^0_L$ events may be extracted with high
confidence for $m_{\pi\pi}$ values between 550 and 730 MeV/c$^2$.  The
confidence in the low $E_{\gamma}$ region, which is deeply sub-threshold, is not
as high, while the situation at the high end of the tagged photon energy range
is similar.

\begin{figure}[hbtp!]
\includegraphics[width=8.3cm]{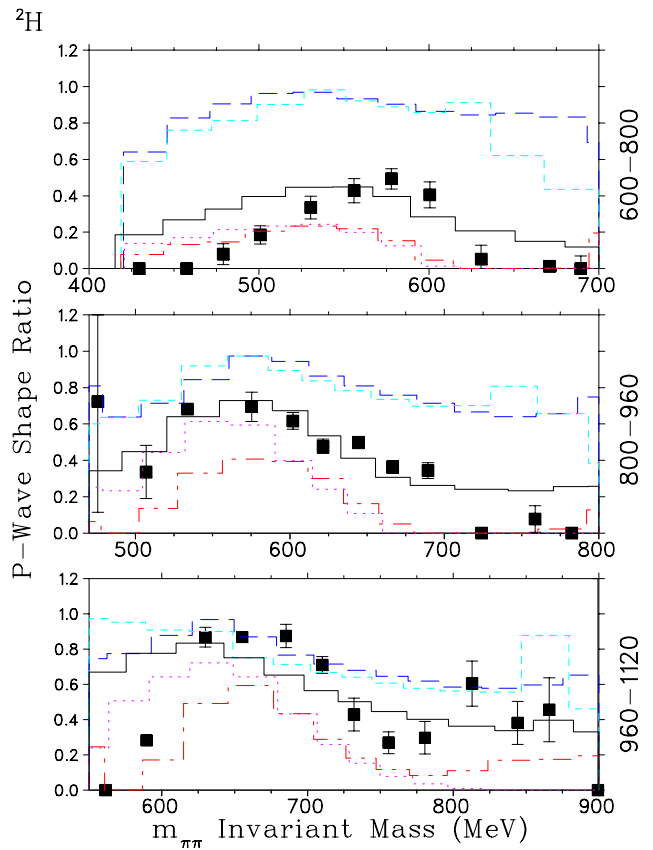}
\caption{\label{fig:comp_mcd} (Color online) $p$-wave shape ratios for the
$^2$H data [$\blacksquare$], quasi-free $\rho^0_L$ simulation [long-dash line],
$N^*(1520)\rightarrow p\rho^0$ simulation [short-dash line], $\Delta^{++}\pi^-$
simulation [dash-dot line] and $\sigma^0$ simulation [dotted line], for the
three tagged photon energy bins, as indicated.  The solid lines are normalized
fits of the simulations to the $R_{p-wave}$ data, as described in the text.
All distributions shown are with $\theta_{\pi\pi}>120^o$ and $1745<m_{miss}\leq
1966$ MeV/c$^2$ (MM090) cuts applied.}
\end{figure}

Fig. \ref{fig:comp_mcd} displays the $p$-wave shape ratio for the $^2$H data
and the equivalent four MC simulations.  In this case, the imposed missing mass
cut is at 90 MeV excitation energy equivalent (MM090), in order to improve the
foreground to background ratio of the data.  The opening angle cut remains at
$120^o$.  The $^2$H data exhibit a $p$-wave ratio which is diluted in
comparison to the pure-$\rho^0_L$ channels, which indicates that a portion of
the data remaining after the cuts is due to background processes.  This is
particularly true for the low-energy panel, where the 550-620 MeV/c$^2$
invariant mass data have a $\sim 40\%$ ratio, while the pure-$\rho^0_L$
channels have ratios in excess of 80\%.  In this case, the dilution of the
ratio could be explained by attributing approximately 40\% of the data to
background processes, while the background proportion would be much less for
the other two energy bins.

\begin{figure*}[hbtp!]
\includegraphics[height=16.cm,angle=90.]{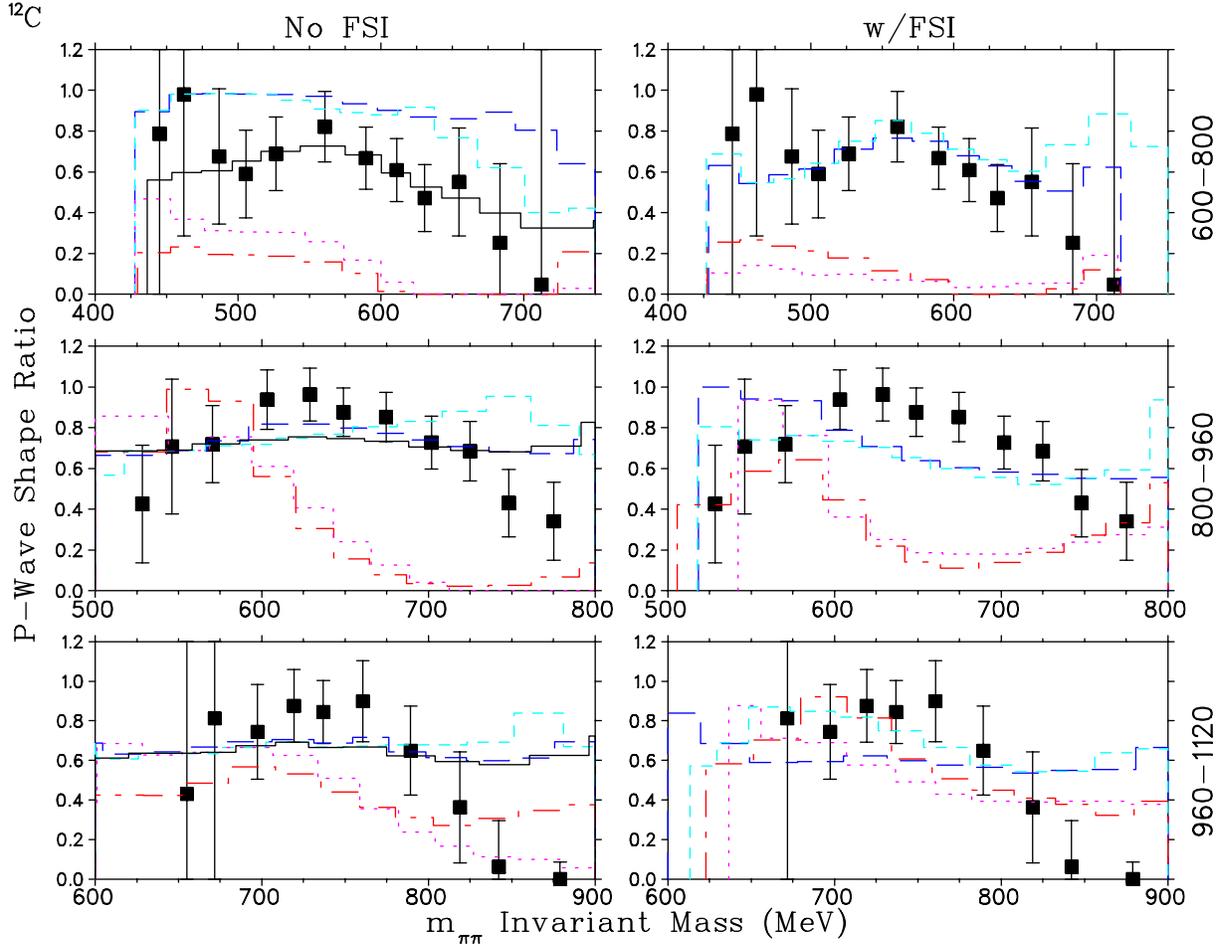}
\caption{\label{fig:comp_mcc} (Color online) $P$-wave shape ratios for the
$^{12}$C data [$\blacksquare$], quasi-free $\rho^0_L$ simulation [long-dash
line], $N^*(1520)\rightarrow p\rho^0$ simulation [short-dash line],
$\Delta^{++}\pi^-$ simulation [dash-dot line] and $\sigma^0$ simulation [dotted
line], for the three tagged photon energy bins, as indicated.  The panels on
the left assume no re-scattering final state interactions, while those on the
right assume that one of the two detected pions has re-scattered.  The solid
lines are normalized fits of the simulations to the $R_{p-wave}$ data, as
described in the text.  All distributions shown are with
$\theta_{\pi\pi}>120^o$ and $10970<m_{miss}\leq 11305$ MeV/c$^2$ (MM130) cuts
applied.}
\end{figure*}

Finally, a measure of the reliability of the helicity analysis for $^{12}$C can
be pursued with the same methodology in Fig. \ref{fig:comp_mcc}.  The data
display larger $p$-wave shape ratios than either of the $^2$H or $^3$He data,
consistent with a smaller proportion of background reactions surviving the two
cuts.

Simulations both with and without the effects of re-scattering FSI are also
displayed in Fig. \ref{fig:comp_mcc}.  In the 600-800 MeV photon energy bin,
the data and FSI-modified $\rho^0_L$ processes are in excellent agreement for
all values of $m_{\pi\pi}$.  This is likely a fortuitous agreement, as it would
indicate a 0\% contribution due to background processes in this energy region,
an expectation at variance with our findings for the two lighter nuclei.
Moreover, in the 800-960 MeV energy range, the data display a $p$-wave shape
ratio which is {\em larger} than that expected for the FSI-modified $\rho^0_L$
processes.  To be in agreement with the data ratios, a {\em negative} number of
background process events would have to be subtracted in this region, which is
unreasonable.

Simulations without FSI are shown in the left panels of
Fig. \ref{fig:comp_mcc}.  There, the data show large $p$-wave ratios which are
consistent with or degraded slightly from those expected for pure $\rho^0_L$
processes.  Our expectation from Fig. \ref{fig:cut12CMC880} is that a large
fraction of the re-scattering FSI will be removed by the analysis cuts.  If one
assumes that 100\% of FSI processes have been eliminated, the non-$\rho^0_L$
contribution to the remaining 600-800 MeV and 800-960 MeV tagged photon bin
yield is still less than 10\%.  A more likely scenario is that some FSI
contribution remains after the cuts, but since the re-scattering processes do
not significantly affect the $p$-wave shape ratio, their contributions are
relatively difficult to distinguish.  In this case, however, the assumption of
no re-scattering FSI actually leads to the subtraction of a greater amount of
non-$\rho^0_L$ background, and so is the more conservative of the two choices.
The 960-1120 MeV energy bin is less definitive, due to poorer statistics.  Even
here, the data exhibit $p$-wave ratios consistent with $\rho^0_L$ dominance,
although the background contribution is likely higher than in the two lower
tagged photon energy bins.

It is interesting to note that among the three nuclei, the more massive and
complex target exhibits the cleanest $p$-wave signature overall.  This analysis
simply underscores the observations made by comparing
Figs. \ref{fig:3hemmthcut}, \ref{fig:dmmthcut}, and \ref{fig:cmmthcut}, which
are $m_{\pi\pi}$-integrated distributions for the data surviving the cuts from
each nucleus.  It appears that the missing mass cut is very effective in
eliminating background, moreso in the case of $^{12}$C.

The response of the MC simulations to the two cuts provides a good description
of the response exhibited by the experimental data.  While all three nuclei
seem to be dominated by the same production mechanisms, as indicated by the
distributions of several important observables, and all three show clear
signatures of $p$-wave attributed to $\rho^0_L$ production and decay, the
relative background proportion is different for the three nuclei.  It is
important to emphasize here that the information obtained for the background
channels are relative ratios, since absolute cross sections are not known under
these conditions.  However, they are valuable inputs to the robustness of the
conclusion that $\rho^0_L$ events are correctly identified for specific regions
of photon energy and invariant mass.  They also provide an additional, and
independent, tool to compare the results of this analysis with previous
publications on the subject \cite{Lo98,Hu98,Ka99} by forming the foundation
upon which the extraction of the $\rho^0_L$ invariant mass distribution is
accomplished.  For all three nuclei, the 800-960 MeV tagged energy regions have
the best statistics and cleanest $p$-wave signatures associated with $l=1,\
m=0$ quantum states.  This is also the energy region of increasing, but still
sub-threshold, $\rho^0$ production.  Thus, this is the region where one expects
non-coherent production signatures to manifest themselves, if they are to be
present at all.  Finally, the analysis and FSI simulations shown here have not
altered the conclusions in Ref. \cite{Lo02}, that no background processes have
been identified which can account for or explain the helicity signatures
observed in the data.

\subsection{Non-$\rho^0_L$ background subtraction}

Based on the analysis of the $p$-wave shape ratios of the different
contributing processes, the portion due to non-$\rho^0_L$ background can be
estimated and subtracted.  The basic procedure was that relative contributions
of the foreground processes ($N^*(1520)\rightarrow p\rho^0$ and quasi-free
$\rho^0_L$) and background processes ($\Delta^{++}\pi^-$ and $\sigma^0$) were
fit to the $R_{p-wave}$ distributions of the data, yielding two relative
normalization factors $\eta_{\rho}$ for the foreground and $(1-\eta_{\rho})$
for the background, respectively.  Our estimate of the in-medium $\rho^0_L$
invariant mass distribution is then obtained from
\begin{displaymath}
\rho^0_L(m_{\pi\pi},E_{\gamma}) Distribution = C_{data} - (1-\eta_{\rho}) C_{background}.
\end{displaymath}
A single normalization factor was thus obtained for each tagged photon energy
bin; the values are tabulated in Table \ref{tab:eta_rho}.  To minimize model
sensitivity, the $A$, $B$, $C$ coefficients for the two foreground processes,
quasi-free $\rho^0_L$ and $N^*(1520)\rightarrow p\rho^0$, and for the two
assumed background processes, quasi-free $\Delta^{++}\pi^-$ and $\sigma^0$,
were averaged together prior to fitting the $R_{p-wave}$ distributions.

\begin{table}[hbtp!]
\caption{\label{tab:eta_rho}
Relative proportion of foreground processes (quasi-free $\rho^0_L$ and
$N^*(1520)\rightarrow p\rho^0_L$) in the data, after missing mass and opening
angle cuts applied, as determined from fitting the $R_{p-wave}$ distributions
of the data.}
\begin{tabular}{c|ccc}
Nucleus   & \multicolumn{3}{c}{Tagged photon energy bin (MeV)} \\
          & 600-800        & 800-960        & 960-1120 \\ \hline
$^2$H     & $0.66\pm 0.06$ & $0.75\pm 0.04$ & $0.87\pm 0.06$ \\  
$^3$He    & $0.25\pm 0.11$ & $0.91\pm 0.02$ & $0.83\pm 0.05$ \\
$^{12}$C  & $0.90\pm 0.01$ & $0.98\pm 0.02$ & $0.96\pm 0.14$ \\
\end{tabular}
\end{table}

The solid lines in Figs. \ref{fig:comp_mch}, \ref{fig:comp_mcd} and
\ref{fig:comp_mcc} indicate the fits to the $R_{p-wave}$ distributions for the
three target nuclei.  It is clearly seen that the MC simulation sums do not
perfectly describe all of the features of the data.  However, the $p$-wave
shape ratios are relative ratios, independent of the actual yield of the data
within each invariant mass bin, and the discrepancies between the fits of the
simulations and the data are largely confined to the tails of the invariant
mass distributions.  To avoid unwarranted sensitivity to these regions, the
$R_{p-wave}$ fits were weighted according to the data yield within each
invariant mass bin.  In many cases, the fit to the data could have been
improved by choosing one of the background or foreground processes alone in the
fit and excluding the other, rather than averaging them.  If this was done, the
obtained normalization factors would not vary outside the uncertainties listed
in Table \ref{tab:eta_rho}.  Nonetheless, which background or foreground
process to choose to obtain the best fit would depend on target nucleus and
photon energy bin, and so the averaging procedure was maintained to ensure
uniformity of approach for all of the data distributions.

\begin{figure*}[hbtp!]
\includegraphics[height=16.cm,angle=90.]{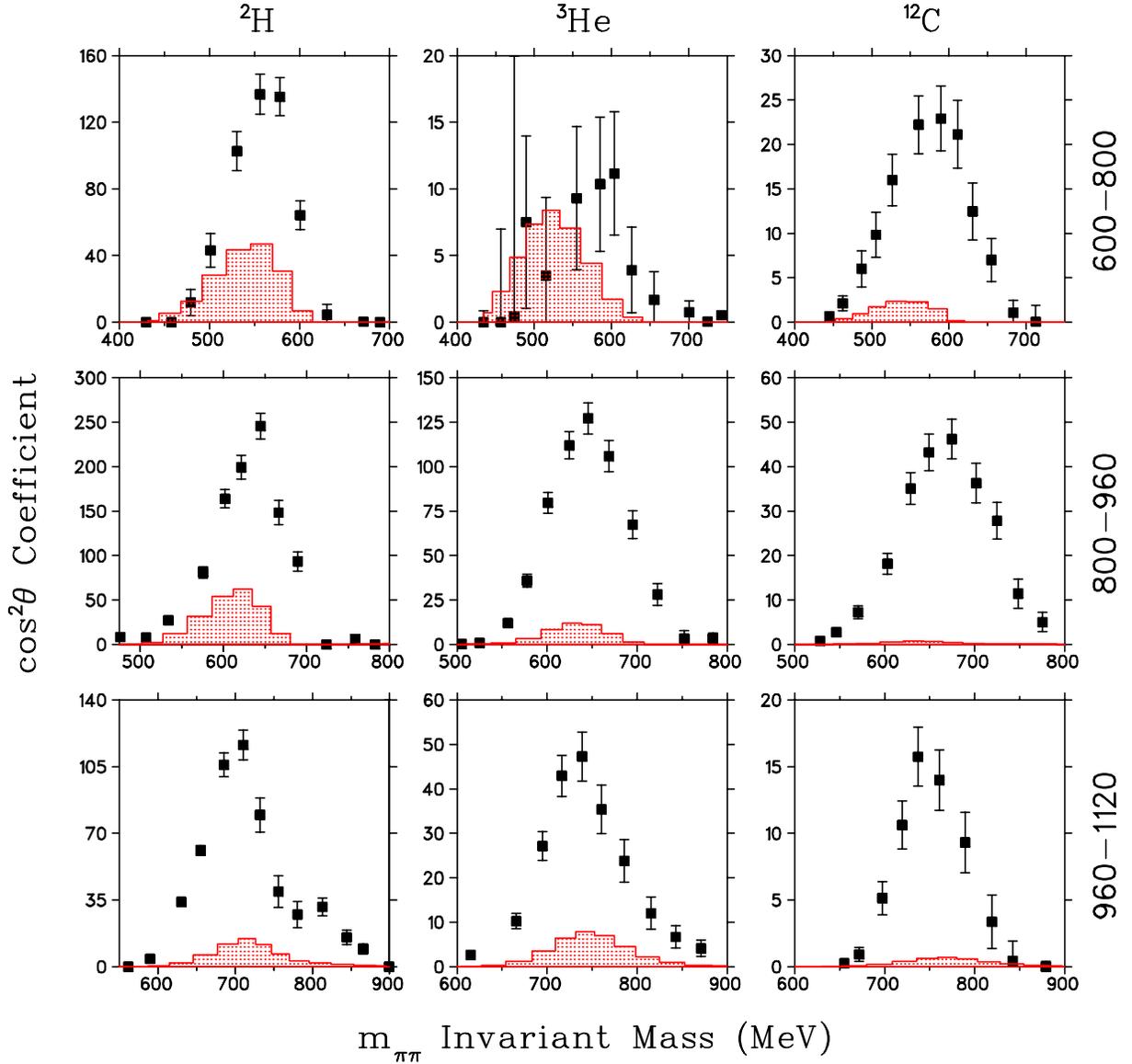}
\caption{\label{fig:cos23} (Color online) cos$^2\theta^*_{\pi^+}$ fit
coefficient `$C$' of the data [$\blacksquare$] and the estimated non-$\rho^0_L$
background contribution [shaded region], based on the fit to the $R_{p-wave}$
distribution.}
\end{figure*}

Fig. \ref{fig:cos23} shows the cos$^2\theta^*_{\pi^+}$ fit coefficient `$C$'
versus invariant mass for all three nuclei.  The shaded regions are for the
estimated background contributions to the cos$^2\theta^*_{\pi^+}$ coefficient
distributions resulting from the $R_{p-wave}$ analysis.  With the exception of
the 600-800 MeV $^3$He distribution, the estimated background contribution is
nearly negligible for $^{12}$C, a bit larger for $^3$He, and larger yet for
$^2$H.  This systematic behavior is despite the tightest missing mass cut being
applied to the $^2$H data, and the most generous cut applied to the $^{12}$C
data.  In most cases, the shaded contributions to be subtracted are relatively
small, quantifying the earlier discussion that the data display large $p$-wave
signatures which are consistent with $\rho^0_L$ decay \cite{Lo02}.

Regarding the 600-800 MeV distribution for $^3$He, one should note from
Fig. \ref{fig:egam_spec} that the photon energy mean and distribution for this
bin are markedly different than for $^2$H and $^{12}$C, and so substantial
differences are to be expected.  Because of the 60 MeV lower mean photon
energy, the $\pi^+\pi^-$ cross-section on $^3$He is reduced, and the data have
poor statistical confidence.  The large data error bars preclude any meaningful
conclusions to be drawn with this analysis technique.  Even with these
limitations, the value of the cos$^2\theta^*_{\pi^+}$ coefficient for
$m_{\pi\pi}>575$ MeV/c$^2$ indicates the presence of some $\rho^0_L$
contribution to the data.  The maximum $R_{p-wave}$ values of the 600-800 MeV
$^2$H data are nearly double those of the background processes, and this can be
accommodated by assuming a foreground/background proportion of 2:1 (Table
\ref{tab:eta_rho}).  The background contribution to be subtracted (shaded
region of the left panel of Fig. \ref{fig:comp_mcd}) has nearly the same
invariant mass distribution as the data, and so the resulting distribution is
not particularly sensitive to errors in the subtracted proportion
normalization.  The 600-800 MeV $^{12}$C data are fitted quite well by the sum
of the MC simulations, and this is reflected in a substantially smaller
uncertainty in $\eta_{\rho}$.

As has been already discussed, the 800-960 MeV region is one of good
statistical precision and one of high confidence in the $p$-wave shape
analysis.  This is reflected in the fits of the simulations to the data, and
the low uncertainties in the $\eta_{\rho}$ normalization factors.  In all
cases, the portion to be subtracted is approximately centered beneath the
data, and so the background subtraction will appreciably alter neither the
centroid nor the width of the extracted in-medium $\rho^0_L$ invariant mass
distribution.  While the $^2$H and $^3$He data are fit quite well, the $^{12}$C
data display a $R_{p-wave}$ ratio which is slightly larger than any of the MC
predictions.  Recall that for the simulations including FSI, this discrepancy
was even larger.  As the maximum $R_{p-wave}$ value of the
$N^*(1520)\rightarrow p\rho^0_L$ simulation equals that of the data, just at a
different invariant mass value, the discrepancy could be in-part due to the
free $\rho^0$ line-shape assumed in the simulations.  The data indicate
a very small non-$\rho^0_L$ proportion to be subtracted, and this is reflected
in the near-unity value of the obtained $\eta_{\rho}$ factor.

The quality of the fits to 960-1120 MeV energy bin data are poorer than at
800-960 MeV.  At this higher energy, $\rho^0\rightarrow\pi^+\pi^-$ decay is
more likely to result in missing mass and opening angle values which do not
pass the physics analysis cuts, resulting in the poorer statistics of the
surviving data sample.  The number of incident tagged photons for this bin also
is smaller, especially for the CD$_2$ experiment.  With the exception of the
extreme tails of the distribution, the fit to the $^2$H $R_{p-wave}$
distribution generally follows the trend of the data, given the size of its
error bars.  For the purpose of estimating the size of the background to be
subtracted, the level of agreement is more than adequate.  The earlier
discussion for the $^{12}$C $R_{p-wave}$ distribution applies here as well; the
data are consistent with a negligible background contribution.

\subsection{Self-consistency checks}

It is of interest to investigate the validity and self-consistency of the
analysis followed above.  Our method is to check whether the fit coefficient
values, used to determine the background contribution to be subtracted, are
able to reproduce the cos$\theta^*_{\pi^+}$ data distributions for a number of
different $m_{\pi\pi}$ bins.  For example, if $R_{p-wave}$ was not a
sufficiently accurate measure of the shape of the data's cos$\theta^*_{\pi}$
distributions, the $A$, $B$, $C$ coefficient values determined from fits to
them would not provide good descriptions of the data, and the validity of the
background subtraction just performed would be in doubt.  The 800-960 MeV
photon energy region was chosen for this check, as it has the best statistical
precision, and so any systematic deviations from the data may be most easily
discerned.

\begin{figure}[hbtp!]
\includegraphics[width=8.6cm]{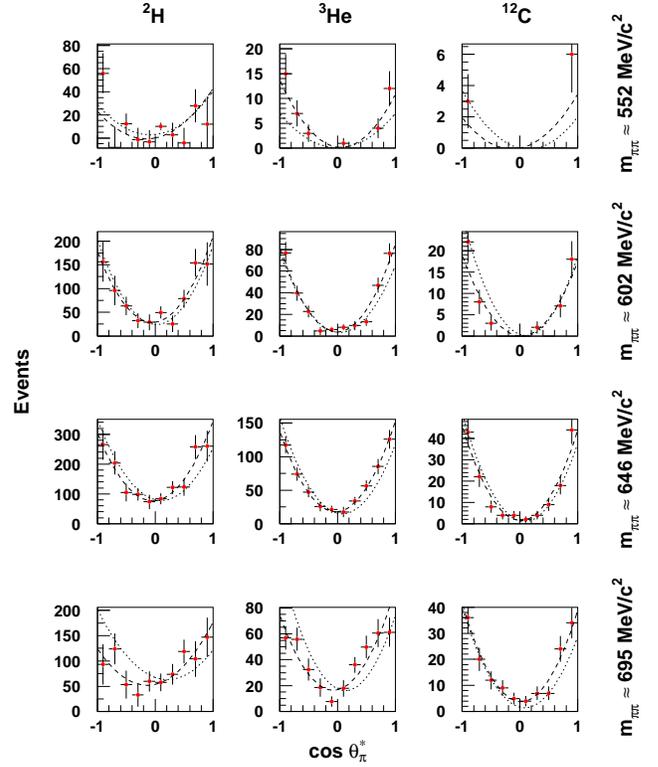}
\caption{\label{fig:cosplot} Helicity angle distributions, with opening
angle and missing mass cuts appropriate for each nucleus applied, for four
invariant mass bins and $800<E_{\gamma}\leq 960$ MeV.  The legend at right
indicates the mean invariant mass value of the data in each bin, averaged over
the three target nuclei.  The long-dashed curves are the fits to the data used
to extract the $R_{p-wave}$ distributions.  The short-dashed curves use the
coefficient values determined from the fits to the $R_{p-wave}$ values
presented in Figs. \ref{fig:comp_mch}, \ref{fig:comp_mcd} and
\ref{fig:comp_mcc}.  The consistency of the two curves, within data error bars,
indicates the validity of the $p$-wave shape ratio analysis.}
\end{figure}

Fig. \ref{fig:cosplot} shows the distributions for four $m_{\pi\pi}$ bins
centered at approximately 552, 602, 646 and 695 MeV/c$^2$.  As discussed
earlier, each invariant mass bin is 90 MeV/c$^2$ wide, and so there is partial
overlap between adjoining bins.  The long-dashed line in each panel is the best
fit to the data with the $A+B$cos$\theta^*_{\pi^+}+C$cos$^2\theta^*_{\pi^+}$
function.  The short-dashed line is a result of the sum of the $\rho^0_L$ and
background contributions, as discussed in the previous section.  The agreement
of the two curves is a direct measure of the reliability of the identification
of the foreground and background components in the data.  For all three nuclei,
the curves agree quite well throughout the $m_{\pi\pi}$ range.  While it is to
be expected that the long-dashed curves will be better descriptions of the
data, as they have been directly fit to them, the deviations of the
short-dashed curves from the data are comparable to the size of the data error
bars.  The 602 and 646 MeV/c$^2$ bin data have the smallest statistical error,
and the two curves are in the best agreement.  The 552 and 695 MeV/c$^2$ bins
have significantly fewer events, and the discrepancy between the curves is
larger, as expected.  The overall conclusion is that there is no systematic
deviation of the short-dashed curves from the data which is outside the
statistical error of the data and analysis, and so the reliability of the
background subtraction method is validated.  The overall self-consistency
achieved for the $^{12}$C data is quite impressive.  This is a direct result of
the very effective suppression of the background processes by the two cuts.

\section{Comparison of the extracted in-medium $\rho^0_L$ invariant mass 
distributions with model calculations}

Before our extracted invariant mass distributions can be compared to any model,
a number of effects must be taken into account in its prediction.  Because of
the sub-threshold nature of the experiment, the role of the limited kinematic
phase-space is significant, and must be accounted for.  In addition, although
every effort has been made to quantify and remove non-$\rho^0_L$ background
contamination from the data, the effect of the TAGX spectrometer acceptance and
detection thresholds still remain.  Similarly, the model prediction should be
subjected to the same $\theta_{\pi\pi}$ and $m_{miss}$ cuts as the experimental
data.  The best way to take these effects into account is to embed the model
calculation within a MC simulation of the experiment, and analyze the simulated
events in the same manner as the data.  Only by following a procedure of this
manner can reliable comparisons be made to the data.

In this section, we will first present a comparison of our data to two
kinematic models utilizing the free $\rho^0$ line-shape obtained from PWA
analysis of the $e^+e^-\rightarrow\pi^+\pi^-$ reaction \cite{Be93}.  This free
line-shape provides a superior description of the low mass $\rho^0$ tail,
significantly better than the standard parameterization of Ref. \cite{pdg}.
This will investigate the roles of kinematic phase-space and Fermi momentum and
will provide the first indication of the reaction mechanism responsible for the
significant longitudinal $\rho^0$ polarization observed here.  Then, we will
compare our data to three phenomenological models of the in-medium $\rho^0_L$.
We believe this will provide the most rigorous comparison to date between model
and observation of the in-medium $\rho^0$ characteristics.  In all cases, the
simulated events were tracked through the simulated TAGX spectrometer and have
all acceptance and threshold effects applied.  The simulated events were then
analyzed in exactly the same manner as the data, and the quadratic fit
coefficient to the cos$^2\theta^*_{\pi^+}$ distribution extracted, just as for
the data.

\subsection{Comparison with the $\gamma N_F \rightarrow \rho^0_L N$ kinematic model}

\begin{figure*}[hbtp!]
\includegraphics[height=14.cm,angle=90.]{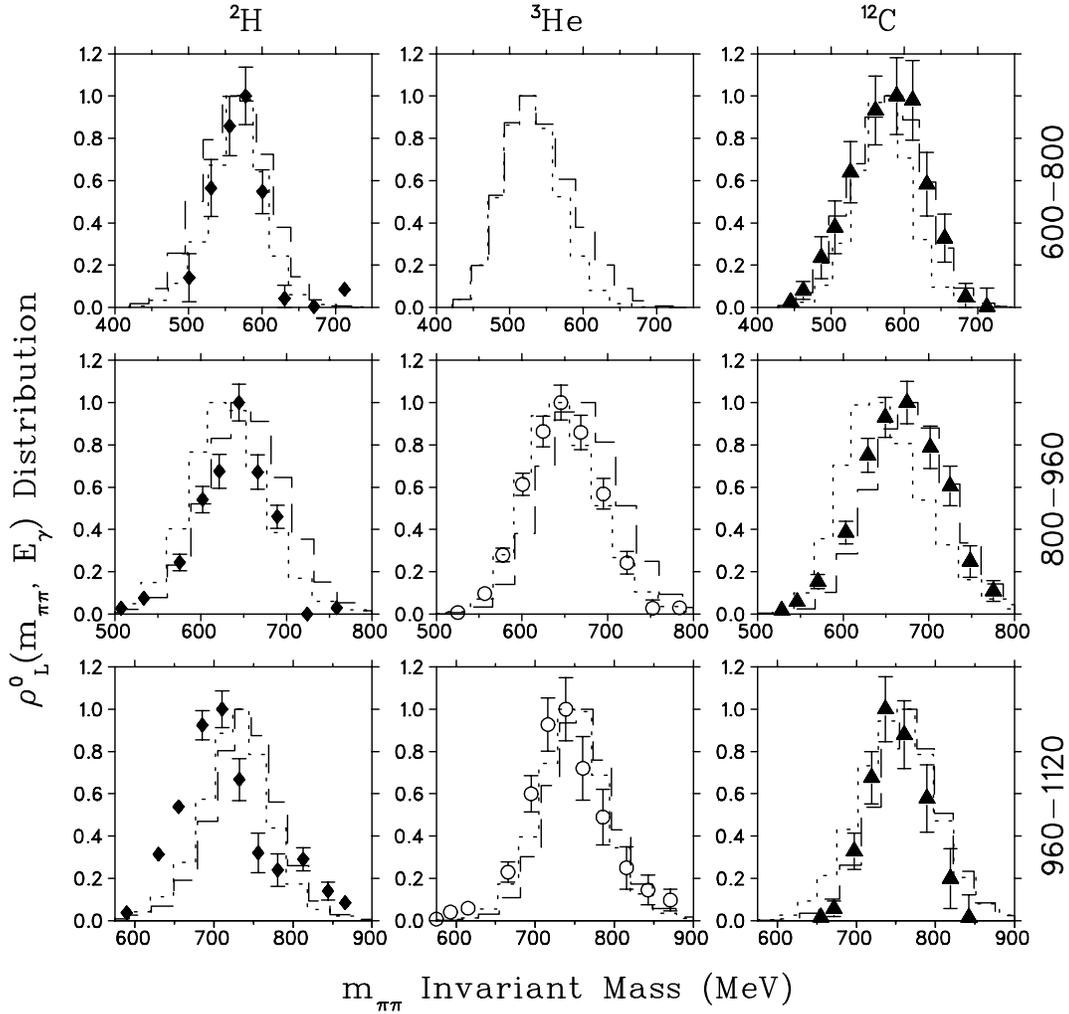}
\caption{\label{fig:mmcomp_rhod133}
Invariant mass distributions of the $\rho^0_L$ strength extracted from the data
[symbols] compared to quasi-free $\rho^0_L$ simulations [long-dash lines] and
quasi-free $N^*(1520)\rightarrow p\rho^0_L$ simulations [short-dash lines],
both incorporating the free $\rho^0$ line-shape from
$e^+e^-\rightarrow\pi^+\pi^-$ partial-wave analysis \cite{Be93}.  The data
error bars reflect both the statistical precision of the data, as well as the
uncertainty in the background subtraction.  To ease comparison, the
distributions have been normalized to a peak amplitude of 1.0.}
\end{figure*}

In Fig. \ref{fig:mmcomp_rhod133}, the extracted invariant mass distributions for
all three nuclei are shown together with the quasi-free $\rho^0_L$ MC simulation
discussed earlier [long-dash lines].  Overall, we see that the simulation
provides a good description of the data for all three nuclei and all tagged
photon energy bins.  This is a convincing statement that indeed the $p$-wave
signatures isolated are those of sub-threshold $\rho^0_L$ production.  The
probability that uncorrelated $\pi^+\pi^-$ pairs from unrelated production
processes could result in an invariant mass distribution similar to the
partial-wave analysis $\rho^0$ and respond in the same manner to the applied
kinematic cuts, is simply too remote to be credible.

Upon closer inspection, we observe deviations between the data and quasi-free
$\rho^0_L$ model distributions.  The agreement between data and simulation is
best for $^{12}$C.  For the 600-800 MeV $E_{\gamma}$ bin, the agreement between
the $^{12}$C MC simulations and the data is excellent for the whole range of
the invariant mass values.  For the 800-960 MeV bin, the width of the simulated
distribution is the same as that of the data, but it is peaked at slightly
higher mass.  In the case of the 960-1120 MeV bin, the quasi-free $\rho^0_L$
distribution is slightly broader than the data and peaked at a higher invariant
mass.  It appears that the trend barely discernible in the 800-960 MeV bin, has
become more pronounced in this energy region.

The same broad conclusions are also true for the comparison between MC and data
for $^3$He.  Here, the widths of the data and MC distributions are nearly the
same, but the data are consistently peaked at about 20 MeV/c$^2$ lower mass.
Surprisingly, the deviations between data and simulation are greatest for
$^2$H.  In both the 600-800 and 800-960 MeV bins, the data distributions are
narrower than those of the simulations, and there is a progressive shift in the
centroids of the data distributions away from the simulations as the photon
energy is increased.

In conclusion, while there is qualitative agreement between the data and the
calculated invariant mass distributions, the deviations merit further study.
The higher centroids and broader widths of the quasi-free MC distributions may
indicate more phase-space available to the assumed production and decay
reaction than the data justify.

\subsection{Comparison with the $\gamma N_F \rightarrow N^*(1520) \rightarrow 
\rho^0_L N$ kinematic model}

Fig. \ref{fig:mmcomp_rhod133} also compares the data distributions to the
quasi-free $N^*(1520)\rightarrow \rho^0_L N$ mechanism MC presented earlier
[short-dash lines].  The assumed free $\rho^0$ line-shape is the same, but the
effects of kinematic phase-space and Fermi momentum are different, due to the
more constrained reaction mechanism.  For the $^2$H 600-800 MeV energy bin, the
agreement between data and the simulation is much improved from the quasi-free
$\rho^0_L$ MC and it is now quite impressive.  The different phase-space
available to the resulting $\rho^0_L$ has narrowed the calculated distribution
and the overall agreement extends throughout the $m_{\pi\pi}$ range.  It is the
strongest evidence yet that $N^*(1520)$ decay is the dominant $\rho^0_L$
production mechanism in this energy range for this nucleus.

The improved agreement between the $N^*(1520)$-based production mechanism and
the $^2$H data carries over to the 800-960 MeV energy range.  In comparison to
the $\rho^0_L$ MC, the simulated distribution has a more similar width to the
data, although the centroids disagree.  For the 960-1120 MeV energy bin, the
data and $N^*(1520)$ MC distributions appear shifted by approximately 40
MeV/c$^2$ from each other, with the data being lower than the model
distribution.  The precise reason for this is not clear but it may be related
to the quasi-free $N^*(1520)$ assumption, admixture of coherently produced
$\rho^0$ and/or admixture of quasi-free $\rho^0_L$ production.  We conjecture
that this admixture may also be responsible for the small bump near
$m_{\pi\pi}=800$ MeV/c$^2$ in this energy bin.

Now moving on to the $^3$He distributions, we see that the $N^*(1520)$
production mechanism gives a much better description of the 800-960 MeV data
than the quasi-free $\rho^0_L$ simulation.  As expected from the experience
with $^2$H, the $N^*(1520)$-based distributions are both narrower and shifted
to lower values compared to the $\rho^0_L$ quasi-free simulation.  For the
960-1120 MeV energy region, the $^3$He data are also in better agreement with
the $N^*(1520)$ process, but the level of agreement is less impressive than for
the lower energy bin.

Finally, for $^{12}$C the level of agreement with the $N^*(1520)$-based
production mechanism is worse than with the quasi-free $\rho^0_L$ mechanism for
the 600-800 and 800-960 MeV bins.  For the 960-1120 MeV bin, the centroid of
the distribution is given better by the $N^*(1520)$ simulation, but the width
is given better by the quasi-free $\rho^0_L$ simulation.  While the two lighter
nuclei are broadly consistent with dominance of the $N^*(1520)$ channel, for
$^{12}$C, only in the 960-1120 MeV energy regime can some evidence for the
$N^*(1520)$ channel be found.

\subsection{Conclusions from comparison to kinematic models}

The general agreement of the experimental $\rho^0_L$ invariant mass
distributions with the expectations from the PWA-based parameterization verify
that the analysis has indeed identified pions originating from $\rho^0$ decay.
The small Fermi momentum distribution of a single nucleon in $^2$H allows
sub-threshold production of the $\rho^0$, while the small nuclear density is
expected to result in minimal, if any, medium modifications.  As such, $^2$H is
an ideal nucleus to use as proof of principle of the analysis that indeed the
$\rho^0$ has been identified and been isolated in the event sample.  This
analysis, then, should lay to rest any question whether the $\rho^0$ has
been indeed identified at sub-threshold energies.

The $^2$H and $^3$He data invariant mass distributions favor $\rho^0_L$
production via the quasi-free $N^*(1520)\rightarrow\rho^0_L N$ mechanism.  The
two quasi-free mechanisms investigated here have invariant mass distributions
that are significantly different from each other, allowing these comparisons to
be meaningful.  The comparison with the kinematic models has thus given a
strong experimental indication of the likely $\rho^0_L$ production mechanism at
these photon energies.  Regarding the lack of evidence in the $^{12}$C data to
support even a modest content of the latter mechanism, we note that the role
the $N^*(1520)$ plays in nuclei, and its in-medium mass and width, is an
ongoing debate.  Our simulation made use of free $N^*(1520)$ parameters from
Ref. \cite{pdg}.  While there is good evidence to suggest that the $N^*(1520)$
is significantly broadened in complex nuclei, more recent results from
photo-absorption studies on carbon and more massive nuclei shed new light on
this subject \cite{Kr01}.  Even though the photon energies did not exceed 800
MeV in that work, its primary subject was the $N^*(1520)$ excitation and its
possible medium modifications, including broadening, in nuclei.  The lack of
experimental signatures of the resonance in total photo-absorption cross
sections in nuclei like carbon, thus remains a not-understood phenomenon.

Finally, a detailed comparison of the data with the simulation results indicate
deviations from the simple kinematic models considered here, which are likely
due to higher $N^*$ and $\Delta$ resonances and/or reaction mechanism admixtures.
The comparison of the data with the phenomenological models to follow may
shed further light on these discrepancies.  Of the three models presented, two
are based on established hadronic interactions that do not involve quark or QCD
elements.  While both models share a similar philosophy and share some common
elements, there are substantial differences between them.  The third model is
also phenomenological in nature, but it is based on quark degrees of freedom
and its foundations are thus completely different from the other two.

\subsection{Comparison with the Rapp - Chanfray - Wambach (RCW) Model} 

This model \cite{Ra97} has two main contributions to the in-medium spectral
shape of the $\rho$.  First, the interactions of the pions with the surrounding
nucleons and $\Delta$'s accumulate substantial strength in the $\rho$ spectral
function for the lower regions of the invariant mass.  This renormalization of
the pion propagation in a $\pi N\Delta$ gas was first proposed in \cite{Ch93,
Ch96}.  Second, the contributions to the $\rho$ spectral shape from in-medium
$\rho$-baryon scattering are evaluated.  This model reproduces the experimental
data on $p$-wave $\pi\pi$ scattering in free space as well as the pion
electromagnetic form factor in the time-like region.

The first element of the model accounts for medium modifications of the $\rho$
by considering the interaction of the intermediate two pion states, in other
words, the $\rho$ is ``dressed'' by the two pion intermediate state.  The
single pion self energy in these intermediate states is evaluated within the
particle-hole excitations at finite temperature.  Specifically, the pions
interact with the surrounding nucleons and with the thermally excited $\Delta$
states through excitations of the type $NN^{-1}$, $\Delta N^{-1}$,
$\Delta\Delta^{-1}$ and $N \Delta^{-1}$.  Evaluation of these type of
interactions lead to substantial broadening of the $\rho$ spectral function.
The peak is shifted to slightly higher values for the invariant mass, however,
significant strength is added to the low mass region below 600 MeV/c$^2$.

The second element of the model treats the scattering of the $\rho$ in-medium.
The sizable strength of the $\rho NN$ and $\rho N \Delta$ coupling constants,
and the large branching ratios of $N(1720)$ and $\Delta(1905)$ decays to $\rho
N$ final states, led the authors to consider $\rho N(1720)N^{-1}$ and
$\rho\Delta(1905) N^{-1}$ particle-hole-like states called ``rhosobars''.  Such
states have been extended to include the $\Delta(1232)$ resonance.  The
calculated contributions of the overlapping states of such ``rhosobars'' result
in an appreciable enhancement of the invariant mass distribution below 600
MeV/c$^2$ while, at the same time, leading to a depletion of the $\rho$
invariant mass peak.

These two main sources of $\rho$ medium modifications lead to an enhancement of
the two-pion invariant mass spectrum below the region accounted for by free
$\rho$ and $\omega$ mesons and they appear to account very well for the
experimental data on di-lepton production in high energy heavy ion collisions
\cite{CERES} and \cite{Ra99}.  As such, the model provides an alternate
explanation of the di-lepton spectra to that of the chiral phase transition
\cite{Ca95,Li96}.  Although the model is particularly suited to conditions of
nuclear matter at high densities and temperatures (heavy ion collisions at high
energies), it can also find application to nuclear matter at lower densities
such as those found in the nuclear core.  It is a non-relativistic model and
this is the main difference from the model of Post - Leupold - Mosel \cite{Po01}.

\begin{figure*}[hbtp!]
\includegraphics[height=14.cm,angle=90.]{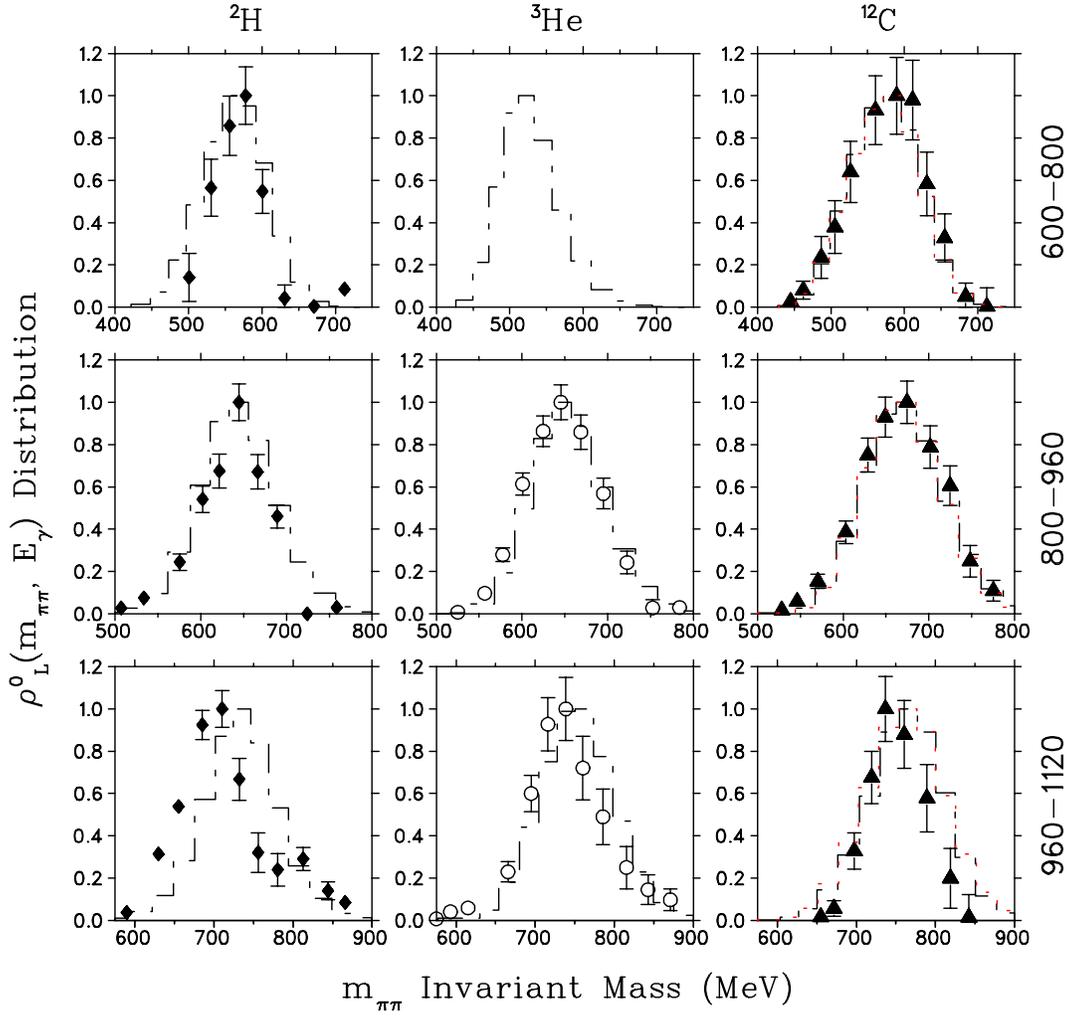}
\caption{\label{fig:mmcomp_rapp3} Invariant mass distributions of the
$\rho^0_L$ strength extracted from the data [symbols] compared to quasi-free
simulations [short dash-long dash lines] incorporating the
Rapp-Chanfray-Wambach model \cite{Ra97} in-medium $\rho^0_L$ spectral function.
In the panels for $^{12}$C, the dotted lines indicate the result when $\pi$
absorption is added to the simulation.  The data error bars reflect both the
statistical precision of the data, as well as the uncertainty in the background
subtraction.}
\end{figure*}

Details on how the RCW-model was implemented in the TAGX simulation are
discussed in App. \ref{app:rcw}.  The RCW-model simulation is compared to the
data in Fig. \ref{fig:mmcomp_rapp3}.  In comparison to the two kinematic models
in Fig. \ref{fig:mmcomp_rhod133}, the RCW model provides a good description of
the data over the full energy range and for all three nuclei.  Its description
of the $^{12}$C data is similar to the quasi-free $\rho^0_L$ kinematic model
while at the same time its description of the $^3$He data is similar to the
$N^*(1520)$ kinematic model.  For the $^2$H data, it does better than the
$\rho^0_L$ quasi-free model, but still not as good as the $N^*(1520)$ model.
This could be due to the limited number of spectral functions provided (see
App. \ref{app:rcw}), which are insufficient to describe the high density
regions of the deuteron.  As with the kinematic models, the agreement with the
data is worst for the 960-1120 MeV bin.  The model curves give broader
distributions than the data justify and this is perhaps indicative of the lack
of the $N^*(1520)$ in the model.  Although $\Delta$'s and $N^*$'s are included,
the latter are of higher mass than the $N^*(1520)$ and play little, if any,
role at our relatively low photon energies.

\subsection{Comparison with the Post - Leupold - Mosel (PLM) Model}

This model is made in the low-density approximation, where the in-medium self
energy of the $\rho$ is completely determined by the $\rho N$ forward
scattering amplitude.  This is a critical element of the model and the $\rho N$
forward scattering amplitudes are extracted from $\pi N \rightarrow\pi\pi N$
processes \cite{Ma92}.  The model is relativistic and it thus avoids the
problems of various components of the theory being evaluated in different
frames of reference, as in other phenomenological models.  In addition, the
problem of the non-coupling of longitudinal $\rho$ states to $p$-wave
resonances in non-relativistic models is resolved in this model.

An essential ingredient of this model is the connection via unitarity between
the matrix element for the decay of a resonance $R$ into a nucleon and a $\rho$
of a given polarization $|M^{L/T}_{RN\rho}|^2$ to the imaginary part of the
forward scattering amplitude $T^{T/L}$: $Im\ T^{T/L} \sim |M^{T/L}_{RN\rho}|$.
All the baryon resonances with sizable couplings to the $\rho N$ channel have
been included with relevant parameters from \cite{Ma92}.  For transversely
polarized $\rho$ mesons, the $N^*(1720)$ resonance plays a dominant role.  In the
longitudinal channel, the $N^*(1520)$ is the main contributor.  For some of the
resonances, there is considerable variation among different parameterizations
of the coupling strength to the $\rho N$ decay channel.

This model concludes that the $N^*(1520)$ plays a large role on the propagation
of $\rho$ mesons in nuclei.  While in the longitudinal channel the differences
between relativistic and non-relativistic calculations are small, in the
transverse channel and at large momenta there are significant differences and
this model leads to pronounced broadening of the $\rho$ invariant mass peak.

\begin{figure*}[hbtp!]
\includegraphics[height=14.cm,angle=90.]{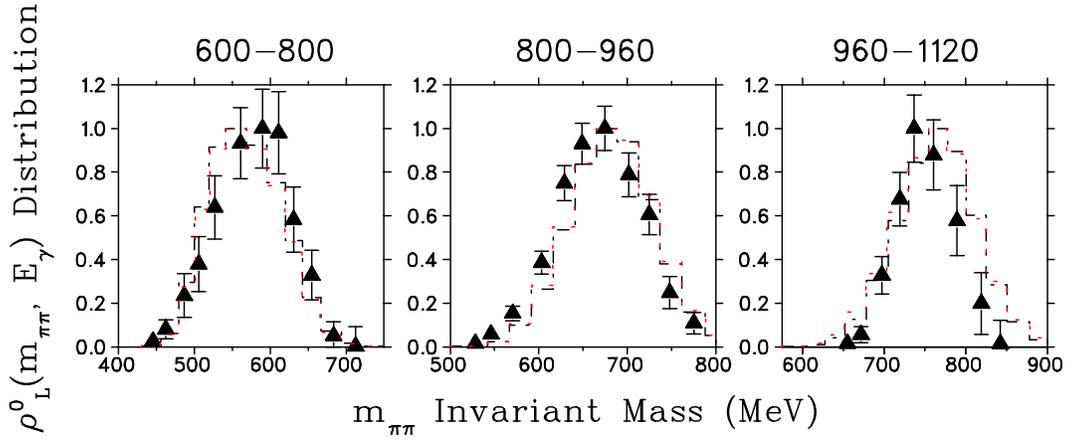}
\caption{\label{fig:mmcomp_post} 
Invariant mass distributions of the $\rho^0_L$ strength extracted from the data
[symbols] compared to quasi-free simulations [dash-dot lines] incorporating the
Post-Leupold-Mosel model \cite{Po01} in-medium $\rho^0_L$ spectral function.
The dotted lines indicate the result when $\pi$ absorption is added to the
simulation.  The data error bars reflect both the statistical precision of the
data, as well as the uncertainty in the background subtraction.}
\end{figure*}

Details on how the PLM-model was implemented in the TAGX simulation are
discussed in App. \ref{app:plm}.  The PLM model authors felt that the model
could only be applied reliably to $^{12}$C, and so a comparison will only be
made to the data for that nucleus, in Fig. \ref{fig:mmcomp_post}.  For the
600-800 MeV tagged photon energy bin, the PLM model provides a good description
of the width of the data distribution, but the centroid is shifted to slightly
lower $m_{\pi\pi}$ values.  No other model has under-estimated the centroid of
this data distribution.  The agreement between the PLM model and the 800-960
MeV data is improved, but is not nearly as good as that provided by the RCW
model.  The agreement with the $E_{\gamma}=960-1120$ MeV data is not as good as
for the mid-energy region, with the model distribution being broader and
shifted to higher invariant mass values.

Nuclear absorption of one or both pions may distort the spectral shape due to
different absorption rates in different density regions of the nucleus, while
medium modifications also depend on nuclear density.  The effect of
pion absorption in the medium was considered via the method described in
section \ref{sec:FSI} and is shown via the dotted curve in the figure.  The
simulation indicates that after the $m_{miss}$ and $\theta_{\pi\pi}$ cuts are
applied, the effect of pion absorption in $^{12}$C is to remove events in an
equal proportion from all parts of the $m_{\pi\pi}$ distribution, resulting in
an invariant mass distribution which is almost indistinguishable from the
simulation in which pion absorption is not considered.

\subsection{Comparison with the Saito - Tsushima - Thomas (STT) Model}

The foundations of this model lie in the quark-meson coupling (QMC-II) version
of the original QMC model of Ref. \cite{Sa95,Sa97} utilizing the MIT quark-bag
model.  Here, quarks in non-overlapping nucleon bags interact self-consistently
with both scalar and vector mesons in the mean-field approximation.  The vector
mesons are themselves described by meson (quark) bags, and the bag parameters
are fixed to reproduce the free nucleon mass and radius.  Since the vector
mesons are described by quark bags as well, additional parameters have been
introduced to fit their free (vacuum) masses.  The in-medium masses are then
given in terms of the mean-field value of the $\sigma$ meson at that density,
which is given by a parameterized (phenomenological) function.

It is worth noting that no pion or $\rho^0$ interactions with the nucleons in
the nuclear medium, $\Delta$ and $N^*$ excitations included, are incorporated
in this model.  The $\rho^0$ interacts with the scalar field of the nucleus and
the two pions emerge with their original four momenta.  Thus, this model
provides a description of the modification of the $\rho^0$ mass due to nuclear
binding effects only.

\begin{figure*}[hbtp!]
\includegraphics[height=14.cm,angle=90.]{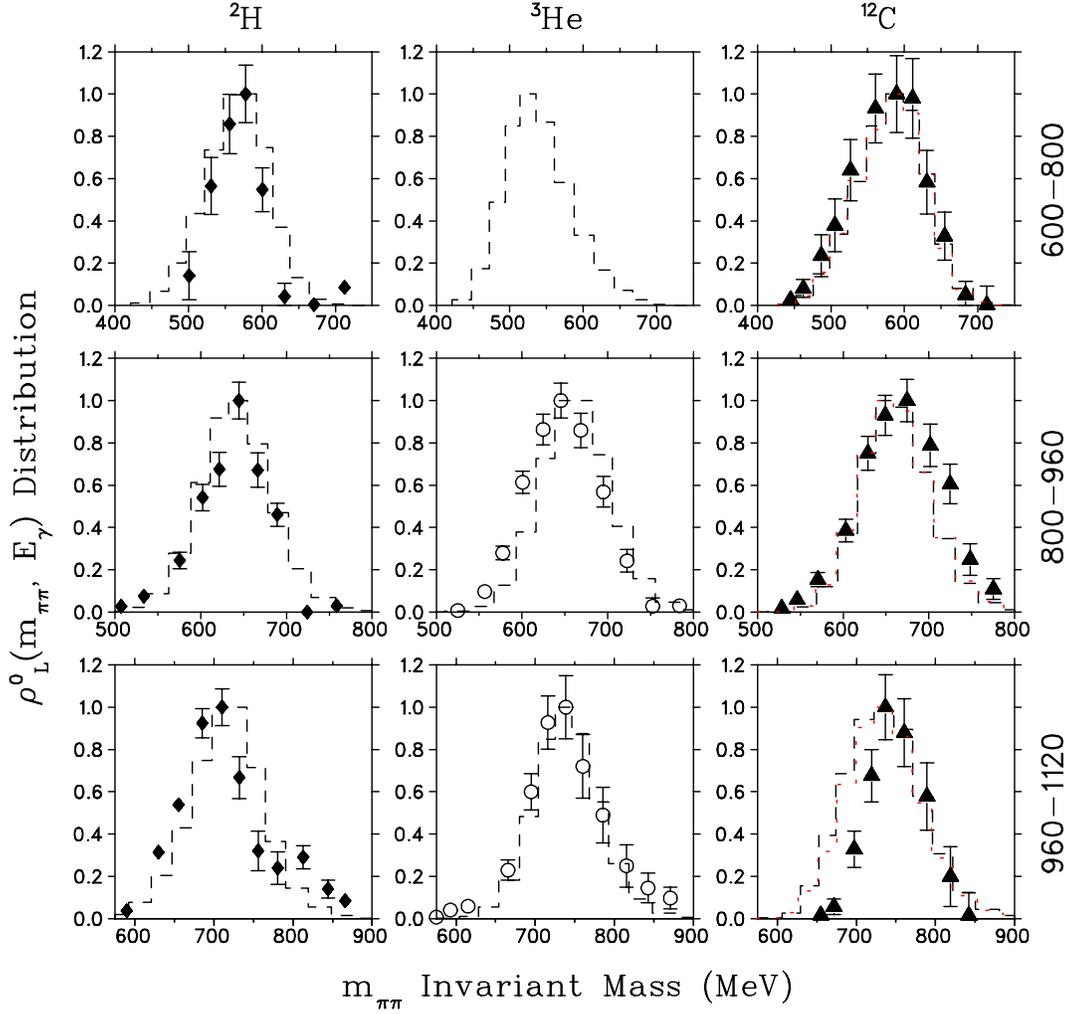}
\caption{\label{fig:mmcomp_saito} Invariant mass distributions of the
$\rho^0_L$ strength extracted from the data [symbols] compared to quasi-free
simulations [dashed lines] incorporating the Saito-Tsushima-Thomas model
\cite{Sa97} in-medium $\rho^0_L$ mass.  In the panels for $^{12}$C, the
dotted lines indicate the result when $\pi$ absorption is added to the
simulation.  The data error bars reflect both the statistical precision of the
data, as well as the uncertainty in the background subtraction.}
\end{figure*}

Details on how the STT model was implemented in the TAGX simulation are
discussed in App. \ref{app:stt}.  The simulation is compared to the data in
Fig. \ref{fig:mmcomp_saito}.  There is a slight shift between the $^2$H and
$^3$He data and model distributions, but overall the agreement is impressive,
especially given the model's lack of any in-medium $\rho^0$ width or shape
modification.  Ref. \cite{Sa97} was the only work to provide specific
predictions for our experiment; their predicted result for $^3$He was a 40 MeV
reduction in the mass of the $\rho^0$.  The mean of the 800-960 MeV data
distribution is lower than the predicted $m^*_{\rho^0}$ value by approximately
20 MeV.  The STT model is the only one to provide a good description of the
$^3$He data distribution at 960-1120 MeV.  As the model has no sensitivity to
spectral shape modifications, this agreement would imply that the $\rho^0_L$
line shape is largely unaffected by the nuclear medium, a fact consistent with
the good agreement with our PWA-based kinematic models.

For $^{12}$C, the agreement between model and data is also quite good.  At
800-960 MeV, the theoretical curve has shifted to slightly lower values
compared to the data, and has a narrower distribution.  This is the only case
where a model distribution is sharper than the data indicate.  This trend
becomes more pronounced in the 960-1120 MeV range and, at the same time, the
distribution has become significantly broader than the data.  This behavior is
unlike all other theoretical or calculated responses and it is unique in its
energy dependence.  It is interesting to note that, in comparison to the
960-1120 MeV data distribution, the STT and RCW-PLM models exhibit shifts in
opposite directions.  A model which incorporates both the nuclear binding
effect of STT as well as the spectral function modifications of RCW-PLM has the
potential to provide the best description of the $^{12}$C data.

\subsection{Conclusions from comparison to phenomenological models}

The first model compared to the data, RCW, provides a reasonable description of
the data from all three nuclei over all energy bins.  Despite the fact that the
$N^*(1520)$ is not explicitly included in the RCW model, the invariant mass
distributions from this simulation exhibit some of the characteristics of the
quasi-free $\rho^0$ and quasi-free $N^*(1520)$ kinematic simulations, presumably
a reflection of the ``rhosobar'' nature of the model.

The second model, PLM, is more sophisticated, as it is fully relativistic.
Where they can be directly compared, the differences between the RCW and PLM
model simulations are small.  While the RCW-simulated distribution for $^{12}$C
tracks the changing shape of the data distribution with energy, the PLM model
predicts a larger energy dependence than the data portray.  The better
agreement between the PLM model and the data at $800< E_{\gamma} \leq 960$
MeV is due, in part, to the fact that it is bracketed by two energy ranges
where the simulated distributions are respectively at too low and too high mass
compared to the data.  Given the inputs and complexity of both models, the
identification of possible factors that allow one model to give a better
description than the other is not an easy task.  However, the clear superiority
of the quasi-free $\rho^0_L$ kinematic simulation over the $N^*(1520)$ for
$^{12}$C may provide a helpful indication of why the RCW model does better.
The RCW model does not include the $N^*(1520)$, while in the PLM longitudinal
channel model it has significant importance.  Since the higher mass resonances
in both models are largely suppressed due to phase-space in this experiment,
we conjecture that the PLM model may over-estimate the role the $N^*(1520)$
plays in $^{12}$C.

Given its simplicity, the STT model does a surprisingly good job of describing
the data.  Because the free $\rho^0$ width is assumed in our simulation, the
good agreement of the STT-simulated distribution with the $^3$He data, for
example, raises the possibility that what is observed for this nucleus is the
combination of a free $\rho^0$ spectral function and a central mass shift.  For
$^{12}$C, the STT model under-predicts the $\rho^0$ mass, and it might seem
reasonable that the combination of the STT and RCW-PLM models will together
provide a good description of the data, i.e. a central mass shift in addition
to a spectral function modification.

\section{Comparison with our previous results}

This work reports new results on $^2$H and $^{12}$C and a re-analysis of our
$^3$He data in light of the helicity analysis reported in Ref. \cite{Lo02}.
Our earlier work on $^3$He \cite{Ka99,Lo98,Hu98,Ma00} reported significant mass
modifications and, in the process, it proved to be controversial.  It is of
interest then to compare the older results with the analysis reported here and
investigate their differences.

The $^3$He analysis in Refs. \cite{Ka99,Lo98,Hu98} was based on the
simultaneous fitting of a number of quasi-free simulations resulting in
$\pi^+\pi^-$ production to data for five different kinematic observables.
Ref. \cite{Hu98} was for the tagged photon energy range of 380-700 MeV, while
Refs. \cite{Ka99,Lo98} reported results for 800-1120 MeV.  In comparison to
this work, these analyses used relatively loose cuts to enhance the $\rho^0$
contribution relative to the background channels.  For example, the analysis of
Ref. \cite{Ka99} was based on a simultaneous fit to the observables from the
entire data sample (no physics cuts) as well as to a set of observables to
which a set of $\rho^0$ enhancement cuts ($70^0< \theta_{\pi\pi} \leq 180^o$
and $2700< m_{miss} \leq 3050$ MeV/c$^2$) were applied.  Because nearly the
entire data sample was used in the analysis, a mass modification result could
even be extracted for the $E_{\gamma}\le 700$ MeV range, while this was not
possible for the more restrictive helicity analysis reported here.  On the
other hand, because the conclusions were explicitly based on the fit of a
series of MC simulations to the data, the results of Ref. \cite{Ka99,Lo98,Hu98}
are more model-dependent than the helicity analysis reported here.  So, the two
independent analyses have their respective strengths and weaknesses, and taken
together they can provide a more complete view of the properties of the in-medium
$\rho^0$ meson.

Refs. \cite{Ka99,Lo98,Hu98} restricted their attention to the `best fit'
$\rho^0$ effective mass value, after accounting for background process
contributions, free-mass $\rho^0$ production, and kinematic phase-space
effects.  Since the emphasis was only on the extraction of the most-probable
reduced-mass value, a Gaussian line shape was taken for the $\rho^0$, using the
free width.  This choice of line shape has been criticized, but the use of a
Lorentzian line shape was investigated in Ref. \cite{Ka99} and yielded almost
identical results.  Even though these two line shapes have quite different low
and high $m_{\pi\pi}$ tails in unresctricted (free) phase space, when the TAGX
thresholds and $E_{\gamma}$ imposed phase space are taken into consideration,
the resulting distributions are nearly identical.  The conclusion of these
studies was that the ``effective'' $\rho^0$ mass in $^3$He was reduced
substantially from its free value.  

Ref. \cite{Hu98} found an effective mass of $490\pm 40$ MeV/c$^2$ for $460<
E_{\gamma} \leq 700$ MeV.  However, no further comparisons to the present
analysis can be pursued.  For the higher energy $^3$He data, Ref. \cite{Lo98}
found the largest improvement in the relative $\chi^2$ fit to the kinematic
distributions at 800-880 MeV, with the next largest improvement for the 880-960
MeV bin.  Ref. \cite{Ka99}, which included several improvements, concluded that
these two energy regimes are the only ones where mass modification could be
reliably extracted, with effective mass values of $m^*_{\rho^0}=642\pm 40$
MeV/c$^2$ for $800< E_{\gamma}\leq 880$ MeV and $m^*_{\rho^0}=669\pm 32$
MeV/c$^2$ for $880< E_{\gamma}\leq 960$ MeV.  The 960-1040 MeV data were
marginally supportive of an effective mass value of $m^*_{\rho^0}=682\pm 56$
MeV/c$^2$, whereas the data in the 1040-1120 MeV region were consistent with no
$\rho^0$ mass modification.  A change in the width of the $\rho^0$ was briefly
investigated in Ref. \cite{Ka99} and it was concluded that {\em ``these
exploratory fits verify the preference for a reduced $\rho^0$ mass, but are
inconclusive, within the sensitivity of the data, as to whether a width
modification is supported in addition.''}  Finally, in Ref. \cite{Pa99} we
suggested that the observed effective $\rho^0$ mass could either be due to the
proximity of the $\rho^0$ to the struck nucleon and the shorter $\rho^0$ mean
decay length at lower energy, or due to the collective excitation of a
$N^*(1520)N^{-1}$ state.

An independent analysis of the $^3$He data was published in Ref. \cite{Ma00}.
Cuts were placed to identify the exclusive $\rho^0$ production ($m_{miss}\sim
m_{^3He}$) and the quasi-free $\rho^0$ production regions
($m_{miss}<(m_{^3He}+m_{\pi})$ for further analysis.  The $\rho^0$ line-shape
was taken as a Breit-Wigner form and non-$\rho^0$ background was modeled via a
phase-space mechanism.  The interferences between free-mass $\rho^0$,
reduced-mass $\rho^0$ and non-$\rho^0$ contributions were taken into account
via the Soding \cite{So66} model.  For the exclusive production region, the
extracted invariant mass distribution was consistent with free $\rho^0$
parameters, presumably these events originated near the periphery of the
nucleus.  The analysis of the quasi-free production region, expected to
correspond to production deeper within the nucleus, yielded a line-shape
significantly at odds with the free $\rho^0$ parameters, implying a
reduced-mass value of $m^*_{\rho}=655\pm 8$ MeV/c$^2$.

For comparison with the present work, we restrict our attention to $^3$He.  The
good agreement of the free PWA line-shape-based simulations with the data
(e.g. Fig. \ref{fig:mmcomp_rhod133}) support the conclusion that the $\rho^0$
width has not been substantially changed from its free value.  This is
consistent with the theoretical expectations of the RCW and PLM models, where
the dominant width modification is restricted to transversely polarized
$\rho^0_T$.  Comparison of the PWA line-shape-based simulations with the
960-1120 MeV data support a mass shift.  Similarly, comparison of the RCW model
\cite{Ra97} with the data (Fig. \ref{fig:mmcomp_rapp3}), indicate that the data
support a slightly lower mass value than the model predicts.  This conclusion
is also reinforced by the comparison of the STT model \cite{Sa97} with the data
(Fig. \ref{fig:mmcomp_saito}), which support a reduction of the $\rho^0$ mass
while using the free width.  The STT model predicts \cite{Sa97} a 40 MeV
reduction in the $\rho^0$ mass in $^3$He, and essentially agrees with the
960-1120 MeV data in Fig. \ref{fig:mmcomp_saito}, confirming $m^*_{\rho}=730$
MeV/c$^2$.  For the 800-960 MeV bin, the STT model distribution is $\sim 20$
MeV high compared to the data, giving a reduced mass value no higher than
$m_{\rho^0}\sim 710$ MeV/c$^2$.  Although these values are obtained with
reference to the STT model, they are consistent with our expectations from the
kinematic and RCW models, and so are listed in Table \ref{tab:comparo} as
representative `effective masses' for this work.  It is difficult to place an
error bar on these masses within the context of the STT model, and so we
refrain from doing so.

\begin{table}[hbtp!]
\caption{\label{tab:comparo} 
$m^*_{\rho}$ results for $^3$He from this work and from Ref. \cite{Ka99}.  For
this work, the masses shown are determined via comparison to the STT model
\cite{Sa97} simulation.  Ref. \cite{Ka99} presented the data contributions from
`free mass' and `reduced mass' $\rho^0$ components separately, and so they must
be combined in a yield-weighted manner before comparison to the results of this
work.}
\begin{tabular}{c|c|c}
              & STT model $m^*_{\rho}$ & Yield-weighted $m^*_{\rho}$ \\
$E_{\gamma}$  & This Work              & Ref. \cite{Ka99} \\ \hline
800-960 MeV   & 700-710 MeV            & $672\pm 31$ MeV \\
960-1120 MeV  & 730 MeV                & $743\pm 17$ MeV\\
\end{tabular}
\end{table}

To compare the old and present analyses, we must not only use the same tagged
photon energy bin limits, but we must also keep in mind that the old
analysis separated the free-mass and reduced-mass $\rho^0$ components, while
the present results are for the total observed $\rho^0_L$ yield.  At 800-960
MeV, free-mass $\rho^0$ production is very small, consistent with the 971 MeV
threshold for $\rho^0$ production one $\sigma$ below nominal mass on the
proton, and then its contribution grows rapidly thereafter.  Yield-weighted
averages of the Ref. \cite{Ka99} results, including the observed free-mass
$\rho^0$ component, are listed in Table \ref{tab:comparo}.  

Given their substantially different analysis techniques, the two sets of
results in Table \ref{tab:comparo} are in broad agreement.  The $m^*_{\rho}$
values from the present work are slightly higher than our earlier results from
Ref. \cite{Ka99}, but agree within 1 $\sigma$.  While the results of the
present work are intrinsically less model-dependent, they rely on tight cuts
upon the data, and so we were unable to extract any result for the 600-800 MeV
energy bin, unlike our earlier works.  This work verifies the conclusions from
Refs. \cite{Ka99,Lo98,Ma00} that there is significant medium modification
observed on the $^3$He target as a result of the sub-threshold production
technique.  A second conclusion regards the issue of $\rho^0$ width
modification.  While our old analysis was unable to make any statement on this
issue, the agreement of the PWA-based simulations with the data distributions
appear to rule out any substantial $\rho^0_L$ width modification, in accordance
with model expectations.

\section{Discussion and Conclusions}

Employing the unique signature of longitudinally polarized $\rho^0_L$ in the
cos$\theta^*_{\pi^+}$ distributions established in Ref. \cite{Lo02}, this work
has identified and isolated $\rho^0_L$ events from tagged photo-production on
$^2$H, $^3$He and $^{12}$C in the $E_{\gamma}=600-1120$ MeV energy region.  The
analysis led to the extraction of the invariant mass distributions of the
$\rho^0_L$ for all three nuclei over three bins of tagged photon energy.  For
$^2$H and $^{12}$C, this is the first time such results are reported.  The
$^3$He data were re-analyzed in the same manner and the results compared with
our previously published results in Refs. \cite{Lo98,Ka99,Ma00}.

The effects of the two cuts used in the helicity analysis have been studied by
comparison to a series of MC simulations with the same cuts applied.  It has
been shown that these cuts cannot create artificial longitudinal helicity
signatures mimicking a $\rho^0_L$ contribution to the data.  A $p$-wave-like
component from non-$\rho^0_L$ residual background was identified with the
assistance of a $p$-wave shape ratio and then subtracted, yielding the
experimental in-medium $\rho^0_L(m_{\pi\pi}, E_{\gamma})$ distributions.  This
background component was found in most cases to be quite small, and its subtraction
yielded $m_{\pi\pi}$ distributions nearly indistinguishable from the
unsubtracted distributions.  Thus, the systematic uncertainty in the extracted
$\rho^0_L$ distributions due to the background subtraction is small.

The 600-960 MeV $^2$H target data favor production via the $\gamma N_F
\rightarrow N^*(1520) \rightarrow \rho^0_L N'$ production mechanism.  For the
$960< E_{\gamma}\leq 1120$ MeV bin, neither of the kinematic production
mechanism models account for the observed distribution.  The agreement of the
data with models based on the PWA-line-shape for the free $\rho^0$ is a
conclusive statement of the identification of the $\rho^0_L$ events in the data
sample, and of the reliability of the analysis.

The $^3$He distributions are also consistent with production via the
$N^*(1520)$ mechanism, over the full observed photon energy range of 800-1120
MeV.  Two phenomenological models \cite{Ra97,Sa97} are quite successful in
describing the data.  Comparison with the STT model simulation, using the free
$\rho^0_L$ width and line-shape indicates that the data provide no evidence of
an in-medium $\rho^0_L$ distribution that is broader than the free
distribution.  This is consistent with model expectations, which predict that
the in-medium $\rho^0_T$ distribution will be significantly wider than the
$\rho^0_L$ distribution.  The STT model \cite{Sa97} predicts a modified mass of
approximately 730 MeV/c$^2$ in $^3$He, and it is in excellent agreement with
the 960-1120 MeV data.  For the 800-960 MeV bin, the data support a somewhat
lower mass, $m^*_{\rho^0_L}=700-710$ MeV/c$^2$.  These results are slightly
higher than the older analyses of the same data in Ref. \cite{Lo98,Ka99,Ma00},
but they confirm the observation of medium mass modifications in the $^3$He
case.

The $^{12}$C distributions are consistent with quasi-free $\rho^0_L$ (PWA)
production over all three tagged photon energy bins.  The data are in excellent
agreement with the RCW model \cite{Ra97} for the two lower energies, but RCW
predicts a wider distribution in the 960-1120 region than the data support.
The PLM model \cite{Po01} does well overall, but results in a larger variation
in the $m_{\pi\pi}$ distribution with energy than the data show.  The STT model
\cite{Sa97} predicts distributions narrower than the $^{12}$C data for the two
lower energy bins and it is the only one of the three phenomenological models
that predicts a lower mass value than the data exhibit.  This indicates that a
model which incorporates both the nuclear binding effect of STT as well as the
spectral function modifications of RCW-PLM has the potential to provide the
best description of the $^{12}$C data.

This very line of argument has already been raised by Brown and Rho \cite{Br02}
in the context of recent results from RHIC.  The Brown-Rho scaling of
medium-dependent masses $m_{\rho}^*/m_{\rho}\sim f^*_{\pi}/f_{\pi}$ is combined
with the spectral function modification of RCW in a unified picture.  In this
case, the $\rho^0$ mass in the phenomenological Lagrangian used by RCW is
replaced with what is essentially the Brown-Rho scaling mass $m^*_{\rho}$.  For
the temperatures and densities appropriate to this work, the two mechanisms
push the $\rho^0$-mass in the same direction, and so result in a lower $\rho^0$
mass than either model alone predict, nearly doubling the density effect
\cite{BrPC}.

The physical mechanism for the dramatic helicity flip remains an open issue.
These data do not support either transversely polarized $\rho^0_T$ or
unpolarized $\rho^0$ production.  Either would have manifested as a
$\sigma^0$-like background associated with some strength at the central region
of the cos$\theta^*_{\pi^+}$ distribution.  In Ref. \cite{Lo02}, one
explanation offered for the observed longitudinally polarized $\rho^0_L$ was
the large $|t|$ in that work associated with $\rho^0$ production
\footnotemark\footnotetext{The data range of $|t|$ values given in
Ref. \cite{Lo02} is incorrect, the actual range displayed by the data is 0.2
to 1.1 (GeV/c)$^2$.}  Rapidly increasing helicity-flip amplitudes had been
observed in \cite{Ba72} for values of $|t|$ larger than 0.4 GeV/c.  An
alternate explanation is that the sub-threshold reaction mechanism may be
responsible for the observed helicity flip.

Finally, the production mechanism seems to depend on the nucleus in question.
Even though $N^*(1520)$ excitation in $^2$H is both broader and quenched
with respect to the free proton, it constitutes a well-identified contribution
to the second resonance group in pion photo-production studies \cite{Kr01} (and
references therein).  For the $E_{\gamma}$ energies of this work, one expects
significant $N^*(1520)$ excitation.  The strong decay channel into $\rho^0$
\cite{pdg} makes this a favorable production mechanism of the low invariant
mass part of the $\rho^0$.  The data are consistent with this expectation.

\section{Acknowledgments}

The authors wish to thank the staff of INS-ES for their hospitality and help
during the experiments, and G.R. Smith for the use of the CD$_2$ targets.  We
also acknowledge the valuable interactions with G.E. Brown, W. Cassing,
M. Post, U. Mosel and R. Rapp.  Their assistance and theoretical insight have
contributed much to this work.  This work has been partially supported by
grants in aid of research by NSERC and INS-ES.

\appendix

\section{The implementation of the phenomenological models in the TAGX
  simulation}

\subsection{The Rapp-Chanfray-Wambach model
\label{app:rcw}}

In-medium $\rho^0$ spectral functions were obtained \cite{RaPC} for cold
nuclear matter at three matter densities, $\rho_B'=\rho_B/\rho_{nuc}=0.5,\ 0.7$
and 1.0, where $\rho_{nuc}$ indicates the standard nuclear density of 0.155
fm$^{-3}$.  Longitudinal, transverse and spin-averaged functions were provided,
but as the data strongly support the production of $\rho^0_L$, only the
longitudinal spectral function was used here.  The in-medium $\rho^0$
propagator for each density is given in matrix form as functions of invariant
mass and three momentum $q$, where $\vec{q}$ is the momentum of the $\rho^0$
with respect to the nuclear medium,
\begin{displaymath}
Im\ D^L_{\rho}=Im\ D^L_{\rho}(\rho_B',m_{\pi\pi},q).
\end{displaymath}
For each simulated event, the tagged photon energy, $\rho^0_L$ mass
$m_{\pi\pi}$, and reaction location within the nucleus were randomly chosen and
an event generated via the quasi-free mechanism $\gamma N_F \rightarrow
\rho^0_L N$, where $N_F$ is the participating struck proton with initial Fermi
momentum $p_F$ and the remainder of the nucleus is a spectator.  It should not
be necessary to consider any other kinematic channel, such as
$N^*\rightarrow\rho^0 N$, as these are already taken into account via the
in-medium spectral function.  To do otherwise would lead to ``double counting''
of resonance effects.  The Fermi momentum distributions of
\cite{Saha,Sch86,VanO} were used.

The nuclear density at the production location was calculated in the following
manner:
\begin{itemize}
\item[$^{12}$C:]{
The electric charge distribution data of Ref. \cite{Ho57} was parameterized as
\begin{displaymath}
\rho (r)=a_0 (1 + a_1 r^2) e^{a_2 r^2}\rho_{nuc}
\end{displaymath}
where $a_0 = 1.2550$, $a_1 = 0.53765$ fm$^{-2}$, $a_2 = -0.39486$ fm$^{-2}$,
and $r$ is in fm.  Note that the standard nuclear density profile formula, used
in Ref. \cite{Sa97}, does not apply for $A<40$.
}
\item[$^3$He:]{
We followed the parameterization given in Ref. \cite{Sa97}, which assumed
$^3$He has a Gaussian density profile with core density of $0.93 \rho_{nuc}$
and $\sigma$ of 1.11 fm.
}
\item[$^2$H:]{
It is perhaps not quite meaningful to consider $^2$H a dense enough nucleus to
expect theoretical models of medium modifications to be applicable.  However,
several theoretical models have all the right ingredients of $\rho$ medium
modifications due to the interactions of pions and resonances with the nucleons
and pions in the nucleus and these models do reproduce the vacuum spectral
shape correctly.  Furthermore, while the average density of the deuteron is
low, recent models utilizing modern $NN$ potentials indicate that the deuteron
is primarily a toroidal structure with a nearly hollow core and a maximum
density of nearly 2 $\rho_{nuc}$ at $r\sim 1$ fm.  Here, the $S$ and $D$-wave
deuteron wave-functions from Ref. \cite{Fo96} using the Argonne AV18 potential
were parameterized and used to calculate the $M_d=0$ state $\rho(r,\theta)$.
}
\end{itemize}
This density information was used two ways.  First, it is used to select the
appropriate spectral function $Im\ D^L_{\rho}(\rho_B')$ for the event.  Second, the
generated event is weighted according to the density at the reaction vertex, so
that production at the diffuse edge of the nucleus is less probable than from
the denser core, consistent with the requirement of a quasi-free reaction
mechanism.

To determine $\vec{q}$, the produced $\rho^0$ is Lorentz-transformed to
the struck proton plus recoil-fragment rest frame, thus specifying the $\rho^0$
propagator value for the event, $Im\ D^L_{\rho}(\rho_B',m_{\pi\pi},q)$.  This value
was incorporated into the statistical weight for the simulated event.  As the
spectral function varies rapidly with $m_{\pi\pi}$ but slowly with
$\rho_B'$ and $q$, the propagator was interpolated from the closest
$m_{\pi\pi}$ bin, but no interpolation was performed either over
$\rho_B'$ or $q$.  The simulated events were then tracked through the
simulated experimental detectors and analyzed in the same manner as the data.

\subsection{The Post-Leupold-Mosel model
\label{app:plm}}

The PLM model was implemented in a manner similar to the RCW model.  In this
case, 15 in-medium propagators were provided \cite{PoPC} in matrix form for
nuclear densities from 0.1 to 1.5 $\rho_{nuc}$.  As with the RCW model,
longitudinal, transverse and spin-averaged functions were provided, but only
the result using the longitudinal spectral function is shown here.  The
propagators were provided as a function of $\rho^0_L$ energy $\omega_{\rho}$ in
the nuclear medium rest frame, as well as $\rho^0$ invariant mass,
\begin{displaymath}
A^L_{\rho}=A^L_{\rho}(\rho_B',m_{\pi\pi},\omega_{\rho}).
\end{displaymath}
and $A^L_{\rho}$ was interpolated from the closest $m_{\pi\pi}$ bin, but no
interpolation was done over $\omega_{\rho}$ or $\rho_B'$.  The same $^{12}$C
density profile and Fermi momentum distributions were used as before.

\subsection{The Saito-Tsushima-Thomas model
\label{app:stt}}

In the STT model, the $\rho^0$ mass is a function of the local density at its
production point.  We use their parameter set $B$ for the mean-field value of
the $\sigma$-meson
\begin{displaymath}
g_{\sigma}\sigma = s_1 \rho_B'+s_2 \rho_B'^2 +s_3 \rho_B'^3
\end{displaymath}
where $s_1=214.0$, $s_2=-44.3$ and $s_3=1.9$ and $g_{\sigma}\sigma$ is in MeV.  The
$\rho^0$ mass is then given by
\begin{displaymath}
m^*_{\rho}=m_{\rho} - \frac{2}{3} (g_{\sigma}\sigma) [1-\frac{8.58\times
10^{-4}({\rm MeV}^{-1})}{2}(g_{\sigma}\sigma)].
\end{displaymath}
The complex manner in which the $\rho^0$ mass and width are incorporated into
the PWA parameterization of Ref. \cite{Be93} preclude its use here, and so the
simpler parameterization of the $\rho^0$ line-shape of Ref. \cite{pdg} is used,
instead.  In all other respects, $\rho^0_L$ are randomly generated within
the nuclear medium via the quasi-free mechanism already described.


\begin{thebibliography}{}
\bibitem{Li99} G.Q. Li, Prog. Part. Nucl. Phys. {\bf 43} (1999) 619-682.
\bibitem{Br91} G.E. Brown and M. Rho, Phys. Rev. Lett. {\bf 66} (1991) 2720.
\bibitem{Ji95} X. Jin and D.B. Leinweber, Phys. Rev. C {\bf 52} (1995) 3344.
\bibitem{Ra97} R. Rapp, G. Chanfray and J. Wambach, Nucl. Phys. {\bf A617} (1997) 472.
\bibitem{Po01} M. Post, S. Leupold and U. Mosel, Nucl. Phys. {\bf A689} (2001) 753.
\bibitem{CERES} CERES Collaboration, Th. Ulrich {\it et al.}, Nucl. Phys. {\bf 
A610} (1996) 317c.
\bibitem{St97} E.J. Stephenson et al., Phys. Rev. Lett. {\bf 78} (1997) 1636.
\bibitem{Oz01} K. Ozawa et al., Phys. Rev. Lett. {\bf 86} (2001) 5019.
\bibitem{bianchi} N. Bianchi {\it et al.}, Phys. Rev. C {\bf 60} (1999) 064617.
\bibitem{Lo98} G.J. Lolos {\it et al.}, Phys. Rev. Lett. {\bf 80} (1998) 241.
\bibitem{Ka99} M.A. Kagarlis {\it et al.}, Phys. Rev. C {\bf 60} (1999) 025203.
\bibitem{Ma00} K. Maruyama, Proceedings of the Second KEK-Tanashi International
Symposium on Hadron and Nuclear Physics with Electromagnetic Probes, Tokyo,
October 1999.  Edited by K. Maruyama and H. Okuno, published by Elsevier
(2000).\\
K. Maruyama, Nucl. Phys. {\bf A 629} (1998) 351c.
\bibitem{Hu98} G.M. Huber, G.J. Lolos, Z. Papandreou, Phys. Rev. Lett. {\bf
80} (1998) 5285.
\bibitem{Lo02} G.J. Lolos {\it et al.}, Phys. Lett. {\bf B528} (2002) 65.
\bibitem{Hu02} G.M. Huber, Proceedings of the Ninth International Conference on
the Structure of Baryons, Newport News, Virginia, March 2002.  Edited by
C. Carlson and B. Mecking, published by World Scientific (2003).
\bibitem{Wa97} D.G. Watts {\it et al.}, Phys. Rev. C {\bf 55} (1997) 1832.
\bibitem{Yoshida} K. Yoshida {\it et al.}, IEEE Trans. Nucl. Sci. {\bf NS-32}
(1985) 2688.
\bibitem{Mar96} K. Maruyama {\it et al.}, Nucl.\ Instr.\ and Meth. 
   {\bf A376} (1996) 335.
\bibitem{tgt3he}  M. Harada {\it et al.}, Nucl. Instrum. Methods
Phys. Res. A {\bf 276} (1989) 451;\\
S. Kato {\it et al.}, {\it ibid.} {\bf 290} (1990) 315;\\
S. Kato {\it et al.}, {\it ibid.} {\bf 307} (1991) 213.
\bibitem{Gar97} G. Garino {\it et al.} Nucl.\ Instr.\ and Meth. {\bf A388}
(1997) 100.
\bibitem{Sh02} A. Shinozaki, Ph.D. thesis, University of Regina, 2002, unpublished.
\bibitem{Est74} P. Estabrooks, A.D. Martin,  Nucl. Phys. {\bf B79} (1974) 301.
\bibitem{Be93} M. Benayoun, {\it et al.}, Z. Phys. {\bf C58} (1993) 31.
\bibitem{Bu96} D.V. Bugg, A.V. Sarantsev, B.S. Zou, Nucl. Phys. {\bf B471} (1996) 59.
\bibitem{Be98} M. Benayoun, {\it et al.}, Eur. Phys. J. C {\bf 2} (1998) 269.
\bibitem{pdg} Particle Data Group, Eur. Phys. J. C {\bf 15} (2000) 1.
\bibitem{Bern81} M. Bernheim {\it et al.}, Nucl. Phys. {\bf A 365} (1981) 349.
\bibitem{Saha} A. Saha and P.E. Ulmer, private communication.
\bibitem{Sch86} R. Schiavilla, V.R. Pandharipande, R.B. Wiringa,
  Nucl. Phys. {\bf A449} (1986) 219.
\bibitem{VanO} J.W. van Orden and P.E. Ulmer, private communication.
\bibitem{Ro78} G. Rowe, M.Salomon, and R.H. Landau, Phys. Rev. C {\bf 18} (1978) 584.
\bibitem{Ash86} D. Ashery, J.P. Schiffer, Ann. Rev. Nucl. Part. Sci {\bf 36} (1986).
\bibitem{Jon93} M.K. Jones, {\it et al.}, Phys. Rev C {\bf 48} (1993) 2800.
\bibitem{geant} GEANT Detector Description and Simulation Tool, CERN
  Laboratory, Geneva, copyright 1993.
\bibitem{Kr01} B. Krusche {\it et al.}, Phys. Rev. Lett. {\bf 86} (2001) 4764.
\bibitem{Ch93} G. Chanfray and P. Schuck, Nucl. Phys. {\bf A555} (1993) 329.
\bibitem{Ch96} G. Chanfray, R. Rapp and J. Wambach, Phys. Rev. Lett. {\bf 76}
  (1996) 368.
\bibitem{Ra99} R. Rapp and J. Wambach, arXiv: hep-ph/9907502 v1.
\bibitem{Ca95} W. Cassing, W. Ehehalt and C.M. Ko, Phys. Lett. {\bf B363} (1995)
  35.
\bibitem{Li96} G.Q. Li, C.M. Ko and G.E. Brown, Nucl. Phys. {\bf A606} (1996) 568.
\bibitem{RaPC} R. Rapp, private communication.
\bibitem{Ho57} R. Hofstadter, Ann. Rev. Nucl. Sci. {\bf 7} (1957) 231.
\bibitem{Sa97} K. Saito, K. Tsushima and A.W. Thomas, Phys. Rev. C {\bf 56} (1997)
 566.
\bibitem{Fo96} J.L. Forest, et al., Phys. Rev. C {\bf 54} (1996) 646.
\bibitem{Ma92} D.M. Manley and E.M. Saleski, Phys. Rev. D {\bf 45} (1992) 4002.
\bibitem{PoPC} M. Post, private communication.
\bibitem{Sa95} K. Saito, A.W. Thomas, Phys. Rev. C {\bf 51} (1995) 2757.
\bibitem{Pa99} Z. Papandreou, G.M. Huber, G.J. Lolos, E.J. Brash,
  B.K. Jennings, Phys. Rev. C {\bf 59} (1999) R1864.
\bibitem{So66} P. Soding, Phys. Lett. {\bf B19} (1966) 702.
\bibitem{Br02} G.E. Brown, M. Rho, arXiv: nucl-th/0206021.
\bibitem{BrPC} G.E. Brown, private communication.
\bibitem{Ba72} J. Ballam {\it et al.}, Phys. Rev. D {\bf 5} (1972) 545.
\end{thebibliography}
\end{document}